\documentclass[twocolumn]{aastex63}

\graphicspath{{./}{Figures/}}

\received{September 6, 2020}
\revised{December 30, 2020}
\accepted{January 3, 2021}
\submitjournal{ApJ}

\shorttitle{Properties of Compact Faint Radio Sources}
\shortauthors{Johnston et al.}
\usepackage{natbib}

\begin{document}

\title{Properties of Compact Faint Radio Sources as a Function of Angular Size from Stacking} 

\correspondingauthor{Jeroen M. Stil}
\email{jstil@ucalgary.ca}

\author{Ryan S. Johnston}
\affiliation{Department of Physics and Astronomy,
University of Calgary,
2500 University Drive NW, 
Calgary, AB, T2N 1N4, Canada}

\author{Jeroen M. Stil}
\affiliation{Department of Physics and Astronomy,
University of Calgary,
2500 University Drive NW, 
Calgary, AB, T2N 1N4, Canada}

\author{Ben W. Keller}
\affiliation{Zentrum f\"ur Astronomie der Universit\"at Heidelberg,
Astronomisches Rechen-Institut,
M\"onchhofstr. 12-14
69120 Heidelberg, Germany}

\begin{abstract}
The polarization properties of radio sources powered by an Active Galactic Nucleus (AGN) have attracted considerable attention because of the significance of magnetic fields in the physics of these sources, their use as probes of plasma along the line of sight, and as a possible contaminant of polarization measurements of the cosmic microwave background. For each of these applications, a better understanding of the statistics of polarization in relation to source characteristics is crucial. In this paper, we derive the median fractional polarization, $\Pi_{0, \rm med}$, of large samples of radio sources with 1.4 GHz flux density $6.6 < S_{1.4} < 70$ mJy, by stacking 1.4 GHz NVSS polarized intensity as a function of angular size derived from the FIRST survey. Five samples with deconvolved mean angular size $1.8\arcsec$ to $8.2\arcsec$ and two samples of symmetric double sources are analyzed. These samples represent most sources smaller than or near the median angular size of the mJy radio source population We find that the median fractional polarization $\Pi_{0,\rm med}$ at 1.4 GHz is a strong function of source angular size $\lesssim 5\arcsec$ and a weak function of angular size for larger sources up to $\sim 8\arcsec$. We interpret our results as depolarization inside the AGN host galaxy and its circumgalactic medium. The curvature of the low-frequency radio spectrum is found to anti-correlate with $\Pi_{0,\rm med}$, a further sign that depolarization is related to the source.        

\end{abstract}


\section{Introduction} \label{sec:intro}
    
    An active galactic nucleus (AGN) often creates jets that travel through the host galaxy and out into intergalactic space to form extended radio lobes filled with relativistic particles, that can extend for hundreds of kpc. The total energy of radio lobes spans a wide range, up to $10^{61}$ erg for the most powerful radio sources \citep{McNamara2005}. Radio lobes displace plasma in the surrounding intracluster medium (ICM) with temperatures averaging around $5\times 10^{7}$ K which, at least temporarily, prevents the gas from settling into a steady state in the gravitational potential of the cluster, a process known as AGN feedback \citep{Fabian_2012}. If a steady state occurred, the gas density in the centre of the cluster would be high enough for the gas to cool and initiate a high star formation rate in galaxies. It is widely suspected that AGN feedback is the mechanism that suppresses such cooling flows \citep{Balogh_2001,Vernaleo_Reynolds_2006,Peterson_Fabian_2006,McNamara2007,Fabian_2012}. For example, AGN can quench star formation by preventing the gas from cooling (thermal feedback), stirring the gas, or by expelling the gas from the galaxy completely (kinetic feedback) \citep{Ishibashi_Fabian_2012}. Since the hot ICM represents most of the baryonic mass of a cluster, AGN feedback can have a profound impact on the formation and evolution of massive galaxies \citep{Silk_Rees_1998,Croton_2006,Nyland_2018}. 

    The radio lobes have high pressure, from both the relativistic electrons and magnetic field, assuming that the energy density of the magnetic field is within an order of magnitude from equipartition \citep[e.g.][]{Kronberg_2004}. The degree of linear polarization of the synchrotron emission is of interest because it reveals ordering of the magnetic field structure on large scales \citep{Burn_1966}. The orientation of a source with respect to the line of sight influences which part of the source is most polarized. This may be due to the projection of the mean magnetic field on the plane of the sky, or a result of the Laing-Garrington effect \citep{Laing_1988,Garrington_1988}.
    The polarization of extra-galactic sources may also be related to other properties of the source structure, including the prominence of jets, hot spots and extended lobes. The accretion mode of the central engine \citep{Osullivan_2017}, and opacity of compact components \citep{Farnes_2014a} have been proposed as physical attributes of the source that affect polarization of the integrated emission. Depolarization by intervening gas-rich systems has been suggested as a contributing factor \citep[e.g.][]{Farnes_2014b}.
   
    Comprehensive analysis of the polarization of bright ($S_\nu \gtrsim 100$ mJy) radio sources has been done by \cite{Mesa_2002} and later by \cite{Tucci_2004} using the NRAO VLA Sky Survey (NVSS) (Condon et al. 1998). \cite{Mesa_2002} found that the median polarization of sources at 1.4 GHz brighter than 80 mJy is $\simeq 2.2$\%, and that fainter sources have a significantly higher percent polarization than for brighter sources. Furthermore, \cite{Mesa_2002} found that the degree of polarization is anti-correlated with the flux density of sources, which was further confirmed by \cite{Tucci_2004} for steep-spectrum radio sources. \cite{Taylor_2007} found the median polarization increases to 4.8\% for sources with flux density of $10 < S_{1.4} < 30$ mJy. \cite{stil_keller_george_taylor_2014} found that from 2 to 20 mJy the median degree of polarization in the NVSS remains $\lesssim$2.5\%, much smaller than previously claimed \citep[][for a review]{stil_keller_george_taylor_2014} but consistent with the findings of \citet{Rudnick_Owen_2014}.

    At higher frequencies, \cite{Sadler_2006} found a median percentage polarization of 2.3\% for a sample of sources selected at 20 GHz, again with a trend for fainter sources to have a higher fractional polarization. \cite{Bonavera_2017} found a mean percentage polarization for AGN sources of $3\%$ to $3.5\%$ from 30 GHz to 353 GHz via stacking Planck satellite data \citep{2018arXiv180706209P}. \cite{Trombetti_2018} found $\sim 3\%$ polarization at 44 GHz and 100 GHz. However, very few sources were detected in polarization at these high frequencies, and their samples are dominated by flat spectrum sources. The available evidence suggests that the median fractional polarization of AGN is remarkably constant from $\sim 1$ GHz to a few hundred GHz.
    
    Several authors have reported a correlation of fractional polarization with angular size in the sense that extended radio sources are more polarized \citep{Cotton_2003,Grant_2010,Rudnick_Owen_2014,Hales_2014,Lamee_2016}. \citet{Vernstrom_2019} examined a sample of 317 physical and 5111 random pairs of bright radio sources and found that the physical pairs were more polarized than the random pairs, while the more widely separated physical pairs were more polarized than the close physical pairs. Correlations with spectral index in the sense that flat-spectrum sources are less polarized \citep{Tucci_2004,2015aska.confE.112S}, may be related as flat spectrum sources are believed to be more compact \citep[]{Sadler_2006}. \citet{Vernstrom_2019} reported a correlation of fractional polarization with 150 MHz to 1400 MHz spectral index with polarization increasing between $\alpha^{150}_{1400} = -0.7$ and $\alpha^{150}_{1400} = -0.5$. Their sample was derived from the catalog of \citet{Taylor_Stil_Sunstrum_2009}, which applied a detection threshold in polarized intensity.
    
    \cite{Rudnick_Owen_2014} presented an ultra-deep polarization survey of the GOODS-N field (resolution of 1.6$\arcsec$, detection threshold of 14.5 $\mathrm{\mu}$Jy beam\textsuperscript{-1}) and found that all sources with detected polarized emission were resolved after convolving the images to a resolution of 10$\arcsec$. In other words, the polarized sources are associated with the small number of sources with emission on angular scales of $10\arcsec$. \cite{Windhorst_2003} derived an angular size - flux density relation of $\Psi(\arcsec) = 2.0\left(S_{1.4}\right)^{0.3}$. For a source with $S_{1.4} \sim 20$ mJy, the median angular size is $\sim$5.0$\arcsec$. This raises the question what sub-population of radio sources is represented in the faint mJy polarized source population, and which plasma is responsible for wavelength-dependent depolarization.
    
    The polarization detections in \cite{Rudnick_Owen_2014} were identified with galaxies in the redshift range $0.2 < z < 1.9$. For a redshift of $0.5 < z < 2.0$, an angular size of 10$\arcsec$ corresponds to a physical size of about 83 kpc in a $\rm\Lambda$CDM cosmology\footnote{Where appropriate, we convert values from older literature to be consistent with this cosmology.} with $\rm H_{0}$ = 67 km $\rm s^{-1}$  $\rm Mpc^{-1}$, $\rm\Omega_{M}=0.3$, $\rm\Omega_{\Lambda}= 0.7$ \citep{2018arXiv180706209P}. 
    
    \citet{Cotton_2003}, \citet{Fanti_2004} and \citet{Rossetti_2008} reported a sudden drop in polarization of Compact Steep Spectrum (CSS) and Gigahertz Peaked Spectrum (GPS) sources \citep[see][for a summary of defining characteristics]{O'Dea_1998} with largest size less than 6 kpc. These authors adopted $H_0 = 100\ \rm km\ s^{-1}\ Mpc^{-1}$ and $\Omega_M = 1$ from the sample of \citet{Fanti_2001}, which means that the corresponding scale in our adopted cosmology is 12 kpc, double the value listed by \citet{Cotton_2003}. \citet{Fanti_2004} found depolarization on scales $\lesssim 2.5$ kpc (5 kpc adjusted for cosmology) at 3.6 cm wavelength and at $\lesssim 4$ kpc (8 kpc adjusted for cosmology) at 6 cm wavelength. This wavelength dependence was attributed to Faraday depolarization \citep{Burn_1966} by the dense interstellar medium in the inner parts of a gas-rich host galaxy \citep{Chiaberge_2009}. In this context, the question whether some of the low-frequency opacity in these sources is attributed to free-free absorption is most interesting. The sub-galactic scale at which this depolarization occurs is much smaller than the $\sim 80\ \rm kpc$ scale suggested by the $10\arcsec$ angular scale found by \citet{Rudnick_Owen_2014}. GPS sources are believed to be the more compact examples of a continuous sequence where larger sources become optically thin at lower frequencies \citep[e.g.][]{O'Dea_1998}. Sources that were historically included in CSS samples tend to display a turnover of the spectrum below $\sim 500\ \rm MHz$. Recently, \citet{Webster_2020} presented a sample of galaxy-scale radio sources identified at 150 MHz that do not have an inverted low-frequency spectrum.
    
    The bright radio source population ($S_{1.4} \geq 100$ mJy) consists mostly of luminous sources \citep{wilman_2008}, brighter than the luminosity boundary between FR I and FR II radio galaxies \citep{fanaroff_1974}. Radio galaxies below the traditional FR I/FR II luminosity boundary dominate the population below $\sim$30 mJy, but recent work has indicated that numerous low-luminosity sources with FR II morphology exist \citep[e.g.][]{Mingo_2019}. \citet{Webster_2020} found 8 times more FR I than FR II in their sample of sources with galaxy-scale jets, although the majority of their sample remained unclassified.

    Very little is known about the polarization properties of this faint radio source population, which may be different from bright sources whose polarization is detectable in current all-sky surveys. Deep surveys of small areas of the sky have revealed some interesting results, but these investigations are limited by small-sample statistics. 
   
    In this paper we use stacking to investigate the relationship between source angular size and polarization of fainter, generally less powerful, AGN. The radio sources investigated here have a total 1.4 GHz flux density between $~6.6 \leq S_{1.4} \leq 70.0$ mJy detected in the NVSS \citep{NVSS} and FIRST \citep{FIRST} surveys. We also derive representative spectral indices\footnote{We assume that $S_{\nu} \sim \nu^{\alpha}$ for spectral index.} for these samples. Stacking is useful because it allows us to investigate much larger samples with a sensitivity similar to that of the deepest small-area surveys. This approach is different from direct observation because it determines the median polarization of a flux density limited sample without detection threshold.
    
    This paper is organized as follows: Section \ref{sec:data_and_methods} describes the method of stacking, the radio survey data used, and the method of sample selection; Section \ref{sec:results} presents the main results of the stacking; and lastly Sections \ref{sec:discussion} and \ref{sec:conclusions} provide a discussion of the results and subsequent conclusions.

\section{Methods and Data} \label{sec:data_and_methods}
   
   \subsection{Stacking} \label{subsec:stacking}
   Stacking is a technique to detect the mean or median signal of a sample of sources if the individual sources, or a significant fraction of the sample, cannot be detected individually. Samples of galaxies selected from optical or infrared surveys, have been stacked at radio wavelengths \citep[][Ocran et al., in prep.]{white_helfand_becker_glikman_vries_2007,Carilli_2008,Garn_2009}. Stacking is particularly effective for investigation of significant samples of rare types of sources, for which it is impractical to do deep observations of each source.
   
   Polarization is a special case for stacking. The target catalog can be selected from the total intensity radio survey, while stacking is performed on the polarization data from the same survey. When combining a large sample of unrelated sources, stacking the Stokes parameters Q and U will result in zero signal because of the random distribution of polarization angles. Stacking is mostly done on images of polarized intensity. This creates additional challenges because the noise in the images is non-Gaussian. The probability density function of polarized intensity, 
   \begin{equation}\label{eq:pol}
        p=\sqrt{Q^2+U^2},
   \end{equation} 
   is given by the Rice distribution \citep{rice1945,vinokur1965},
   \begin{equation}
        F(p\,|\,p_0, \sigma_{QU}) = {p\over \sigma_{QU}} \exp\Bigl[ -{p^2 + p_0^2 \over 2 \sigma_{QU}^2}  \Bigr]J_0\Bigl({i p p_0 \over \sigma_{QU}}\Bigr).
    \label{rice-eq}
    \end{equation}
    Here, $p_0$ is the true polarized signal, $J_{0}$ is the 0\textsuperscript{th} order Bessel function, and $\sigma_{QU}$ is the standard deviation of the noise in the Stokes $Q$ and $U$ images, which is presumed Gaussian. 

    In the limit of no signal, $p_0 = 0$, the Rice distribution reverts to the Rayleigh distribution. For very weak but finite polarized signal, $p_0 < \sigma_{QU}$, the relation between the median of the Rice distribution and the true polarized signal $p_0$ is non-linear. The expectation value of the observed polarized intensity is also much larger than the true signal \citep[e.g.][]{1985A&A...142..100S}. The median observed polarized intensity $p_{\rm med}$ of a sample of sources also depends in a complicated way on the distribution of the true signals \citep{stil_keller_george_taylor_2014}. We return to this issue in Section~\ref{sec:results-stat}.
    
    Stacking gives us the median observed polarized intensity, which is far from the true sample median because of this strong polarization bias. The median polarized signal can be retrieved by means of Monte Carlo Simulations as described by \citet{stil_keller_george_taylor_2014}. Similar procedures were also applied at higher frequencies to polarization from the Planck satellite by \citet{Bonavera_2017} and \citet{Trombetti_2018}. In order to correct for contamination of the cosmic microwave background polarization by AGN, stacking the square of fractional polarization is also applied, with different statistics \citep{Gupta_2019}.
    
    The Monte Carlo simulations aim to reproduce the distribution of observed polarized intensities $p_i$ for a given distribution of true polarized intensities $p_{0,i}$, defined by a distribution function with an adjustable median. The median of the model distribution is then varied until the median of the simulated $p_i$ matches the median of the observed polarized intensities. The shape of the distribution is constrained by the distribution of observed polarized intensities, traced by the ratio of the upper quartile to the lower quartile \citep{stil_keller_george_taylor_2014}. The polarized source counts derived from stacking by \citet{stil_keller_george_taylor_2014} were closely consistent with those derived from the GOODS-N deep field \citep{Rudnick_Owen_2014}, and the implied density of polarized sources has recently been confirmed by early observations of the POSSUM survey.
    
    In this paper we apply Monte Carlo simulations similar to those of \citet{stil_keller_george_taylor_2014}, with a modification for the broad flux density bins of the samples investigated here. The flux density range within a bin is included in the simulations by multiplying the fractional polarization $\Pi_0$ drawn from the assumed distribution with the actual flux density of each source in the sample. This ensures that the range of flux densities in the sample is represented into the simulated distribution of $p_{i}$. A dummy run of the stack on images of Stokes Q or U is made to measure a local noise level $\sigma_{QU,i}$ for each source. Further description of the Monte Carlo simulations is postponed until Section~\ref{sec:results} as it requires details derived during the analysis.
   
    \subsection{The Data} \label{subsec:data} 
    
    In this paper we utilize data from a total of five radio sky surveys. For stacking polarization we use the NRAO VLA Sky Survey\footnote{\url{https://www.cv.nrao.edu}} \citep[NVSS;][]{NVSS}, which covers 82\% of the sky ($\sim33885 \: \rm deg^{2}$) imaged at 1400 MHz in total and polarized intensity. With the VLA in `D' configuration, the NVSS has an angular resolution of $45\arcsec$ (FWHM). However, in this paper we only use the original source catalog for the purpose of sample selection, as discussed in Section \ref{subsec:sample_Selection}. For stacking, we utilize a version of the NVSS constructed by \cite{Taylor_Stil_Sunstrum_2009} where the images were made at $60\arcsec$ resolution. In the original NVSS survey images, intensities are rounded to a fraction of the noise, which interferes with stacking. The NVSS has a root mean square (rms) noise of $\sim 0.45$ mJy $\rm beam^{-1}$ in total intensity and $\sim 0.29$ mJy $\rm beam^{-1}$ in Stokes Q and U. The 90\% confidence range for positions is approximately $\pm 1\arcsec$ in right ascension and declination for strong sources and $\pm 2\arcsec$ for faint sources \citep{NVSS}. These errors are small enough to include position information in our sample selection and interpretation of images from surveys with higher angular resolution in following sections.

    Sample selection of the sources used for stacking involves a combination of the NVSS and the Faint Images of the Radio Sky at Twenty-centimeters survey\footnote{\url{https://sundog.stsci.edu}} \citep[FIRST;][]{FIRST,Helfand_2015}, also imaged at 1400 MHz, but with the VLA in `B' configuration. FIRST covers 25.63\% (10,575 square degrees) of the sky with $5\arcsec$ angular resolution and rms noise of $\sim 0.15$ mJy $\rm beam^{-1}$. Therefore, FIRST has a much better positional accuracy of $<0.5\arcsec$ rms \citep{FIRST} and can distinguish small-scale structures with accurate positions. The largest angular scale detectable by FIRST is $\sim 30\arcsec$.
    
    For spectral analysis we use three additional low-frequency surveys to accompany the higher frequency NVSS, starting with the Westerbork Northern Sky Survey\footnote{\url{https://heasarc.gsfc.nasa.gov}} \citep[WENSS;][]{WENSS}. WENSS covers $3.14$ sr of the sky north of declination $+29\degr$. At 325 MHz, WENSS has a $54\arcsec$ resolution with a positional accuracy of $1.5\arcsec$ and an rms noise of $\sim 4$ mJy $\rm beam^{-1}$. This sensitivity is a good match to the NVSS for steep spectrum sources, but less so for sources with a flat or an inverted spectrum. In this paper we use stacking of total intensity images in all surveys to retrieve a sample median flux density at each frequency.
    
    We also utilized the TIFR GMRT Sky Survey\footnote{\url{http://tgssadr.strw.leidenuniv.nl/}} ADR1 \citep[TGSS;][]{TGSS} made with the Giant Metrewave Radio Telescope (GMRT) which covers $3.6 \pi$ sr of the sky between declination $-53\degr$ and $+90\degr$ imaged at a frequency of 150 MHz. The rms noise of the TGSS is $\langle \sigma \rangle \approx 5$ mJy $\rm beam^{-1}$. The original resolution of the TGSS is $25\arcsec$ but in this work we used the Common Astronomy Software Applications \citep[CASA;][]{2008ASPC..394..623J} task imsmooth to convolve the published TGSS images to $60\arcsec$ resolution.  After convolution, the image noise is $14\ \rm mJy\ beam^{-1}$ rms.
    
    Finally, we use the 74 MHz VLA Low-Frequency Sky Survey\footnote{\url{https://www.cv.nrao.edu/vlss/VLSSpostage.shtml}} \citep[VLSS;][]{VLSS} with a FWHM angular resolution of $80\arcsec$ and an average rms noise of $\sim 100$ mJy $\rm beam^{-1}$. These four surveys have comparable beam-sizes which is ideal for spectral analysis of the integrated emission of radio sources. When combined, these surveys sample the spectral energy distribution from 74 MHz to 1400 MHz.
    
    To gain insight into the morphology of our samples in greater detail, we make use of the quick-look images from the Jansky Very Large Array (VLA) Sky Survey\footnote{\url{https://science.nrao.edu/vlass}} \citep[VLASS;][]{2020PASP..132c5001L}. Similar to the NVSS, VLASS will image the entire sky north of declination $-40\degr$ in Stokes I, Q, and U. But unlike the NVSS and FIRST surveys, VLASS is imaged in the $2$ to $4$ GHz frequency range and will have $2.5\arcsec \times 2.5\arcsec$ (VLA B/BnA configurations) angular resolution with a combined sensitivity of $70 \rm \mu Jy$ rms. The final data products are not yet available. We only do a limited investigation of angular size and morphology using Stokes I quick-look images released through the NRAO website.

    \subsection{Sample Selection} \label{subsec:sample_Selection}

    The sample selection process started with a list of unconfused sources from the unified catalogue of radio sources\footnote{\url{http://www.aoc.nrao.edu/~akimball/radiocat.shtml}} compiled by \cite{Kimball_2008}, and is similar to the sample selection for structure applied by these authors. The catalog combines data of radio sources from four radio catalogs, including the FIRST, NVSS, and WENSS surveys. The data set was reduced to only include sources within the footprint of the FIRST survey, which is completely covered by the NVSS. The FIRST survey specifically avoids the galactic plane ($|b|\lesssim 2\degr$) and was chosen in part to ensure that all sources examined in this paper are extragalactic in nature. This reduced the list from approximately $2.7 \times 10^{6}$ sources to $\sim 1.0 \times 10^{6}$ for further sample selection. From the reduced data set we start by using logarithmic binning to divide the data into 7 flux bins across a NVSS flux density range from $\sim$6.6 to 70.0 mJy, with a bin size of 0.146 dex, or a factor of 1.4 in flux density. 
    
    Sample selection makes use of the recorded sky coordinates of the sources as reported in the NVSS and FIRST catalogs. As a result of the difference in angular resolution of the FIRST and NVSS surveys, independent RA and DEC values in J2000 equatorial coordinates are provided in each survey for the same source on the sky. The NVSS position can be seen as an average of the FIRST positions weighted by the brightness of FIRST sources within the NVSS beam. The NVSS position errors are only $\sim 1\arcsec$ (FWHM), so we can use the angular separation ($r$) between the reported positions (NVSS/FIRST) during sample selection. If we want to select single sources, we can restrict the reduced data-set to only include sources where the reported positions in FIRST and NVSS agree to within a small amount, such as $3\arcsec$. Or if we want to select double sources, we only include sources where the reported positions are different by a significant amount, such as $5\arcsec$ to $7\arcsec$, alongside other selection criteria. 
    
    Along with the restriction on position separation, each flux bin is divided into sub-samples based on two additional parameters. We start with the ratio of the integrated flux densities ($S$) of both NVSS and FIRST. This can be defined as a parameter $\mathcal{R}$ so that
    \begin{equation}\label{eq:flux_ratio}
        \mathcal{R} = \frac{S_{\rm NVSS}}{S_{\rm FIRST}}.
    \end{equation}
    If both surveys measured the same flux density, such that $S_{\rm NVSS}=S_{\rm FIRST}$, then $\mathcal{R}=1$, implying that there is a single source at the centre of the NVSS point spread function. But if $\mathcal{R}=2$ and the reported positions do not align, then FIRST will resolve a likely double source with components of equal brightness within the NVSS beam. This statement will be verified later.
     
    Additionally, we select samples based on the compactness of each source, which is defined as the ratio of the integrated flux density and peak intensity of FIRST. This compactness parameter $\mathcal{C}$ can be written as
    \begin{equation}\label{eq:compactness}
        \mathcal{C} = \frac{S_{\rm FIRST}}{I_{\rm FIRST}}.
    \end{equation}
    The compactness parameter describes how closely related the FIRST peak intensity is to the integrated intensity of the source. $S_{\rm FIRST}$ is derived from fitting a two-dimensional Gaussian to each source. If $\mathcal{C}=1$ then the peak intensity is equal to the integrated intensity and the source is very compact. By extension if $\mathcal{C}>1$ then the source becomes more extended for higher values of $\mathcal{C}$. Later in this paper, we will derive more precise mean angular sizes as a function of $\mathcal{C}$.

    \begin{figure}[h]
        \centering
        \includegraphics[trim={0.25cm 0.325cm 0.25cm 0.25cm},clip,width=\linewidth]{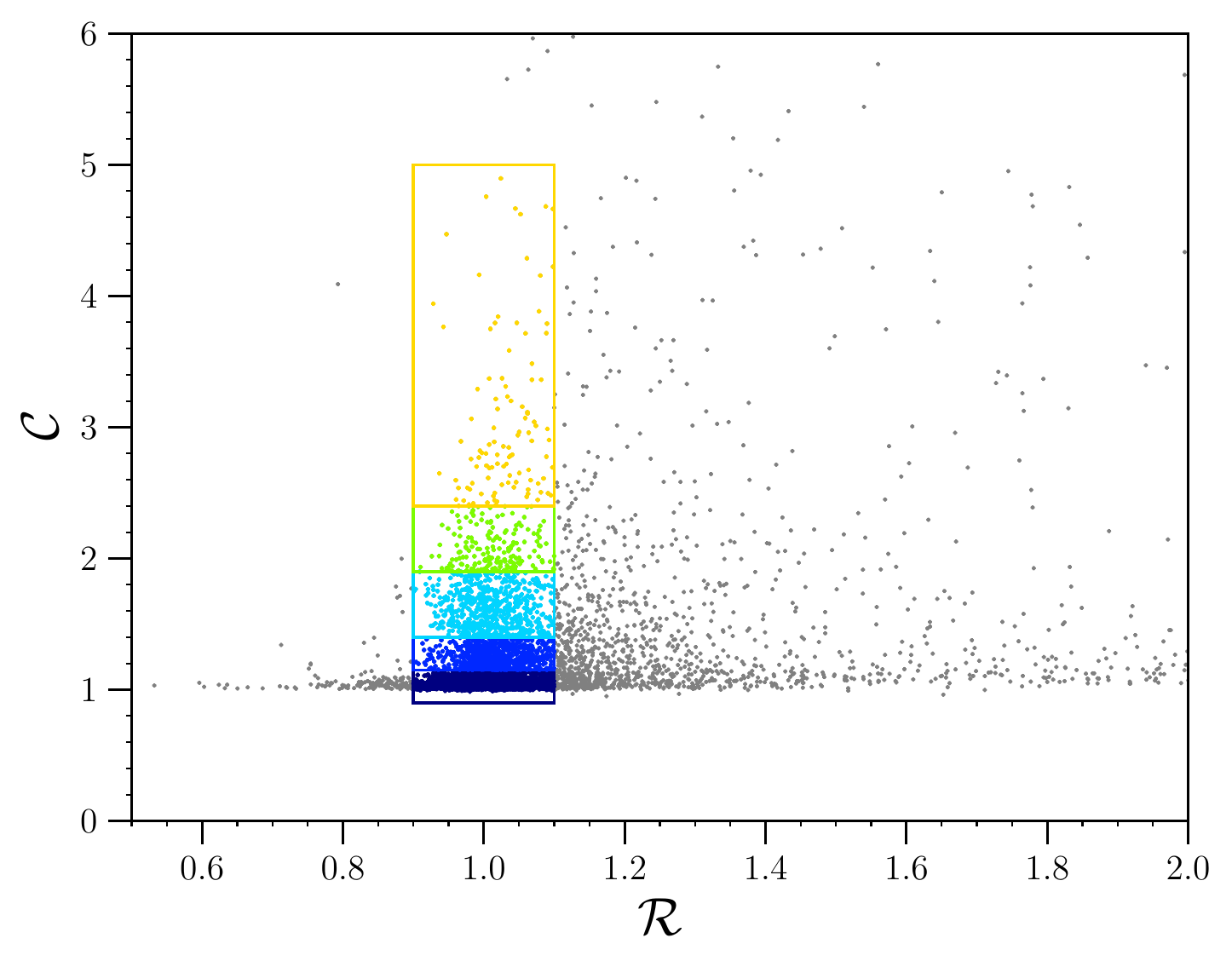}
        \caption{Example of the sample selection process for single sources. The $x$ and $y$ axes represent the parameters derived from the FIRST and NVSS surveys as defined in Equations \ref{eq:flux_ratio} and \ref{eq:compactness}. The sources shown have an NVSS flux density of $50.0\leq S_{\rm NVSS}\leq 70.0$ mJy, with an angular separation of $0 \leq r \leq 3\arcsec$. Here, the dark blue box represents the most compact sources, the blue box represents single sources that are just resolved in FIRST with $\mathcal{C}>1$. The cyan, green, and yellow boxes represent increasingly extended sources. In other words, sources of decreasing compactness. See Figures, \ref{fig:extended_mosaic_1} and \ref{fig:extended_mosaic_2} for examples of the appearance of the sources present in the blue and green boxes. Additionally, see Table \ref{tab:samples_used} for more details on sample the selection parameters.}
        \label{fig:compact_selection}
    \end{figure}
    
    \begin{figure}[h]
        \centering
        \includegraphics[trim={0.25cm 0.325cm 0.25cm 0.25cm},clip,width=\linewidth]{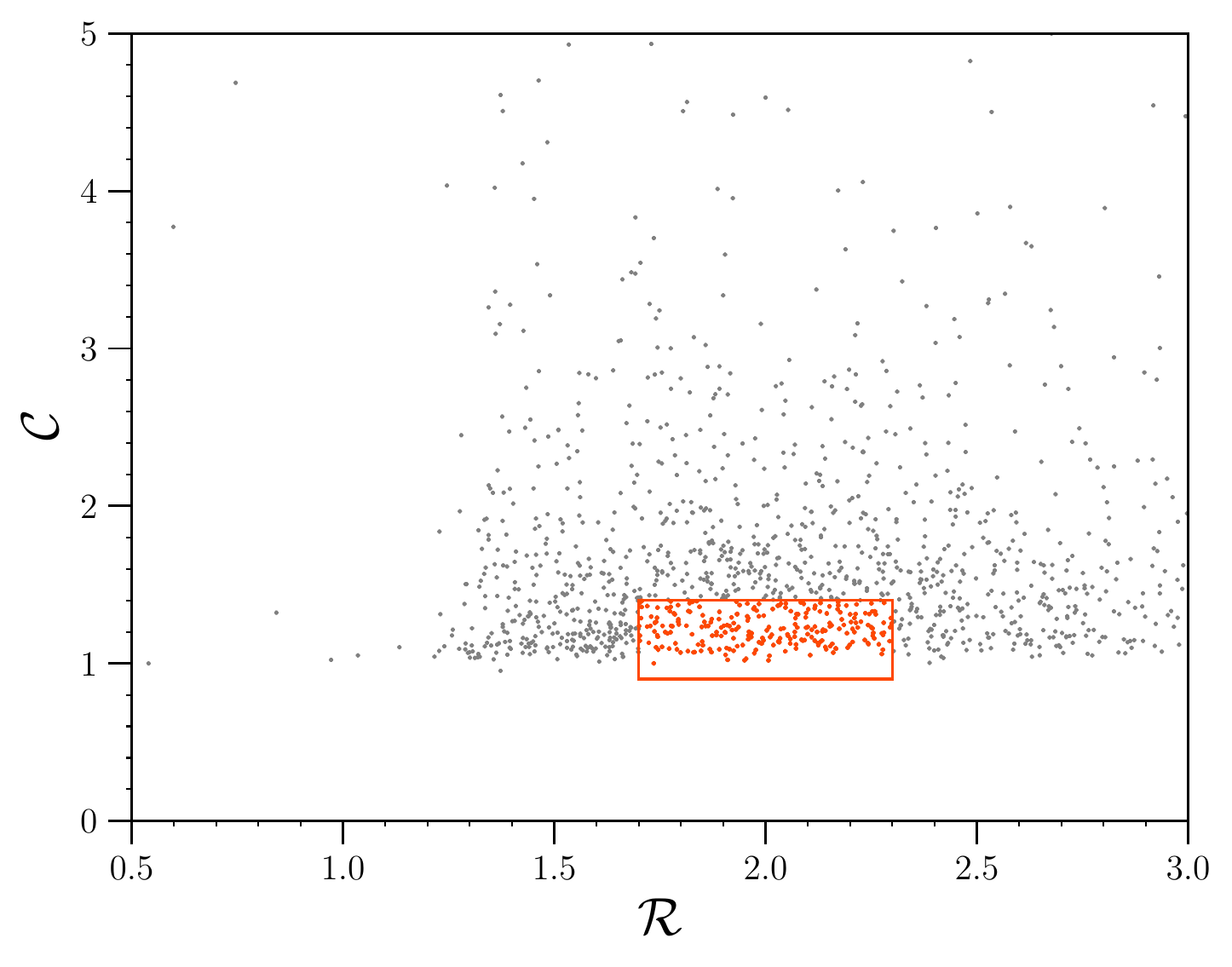}
        \caption{Example of the sample selection process for double sources. The sources are from the same NVSS flux bin as in Figure \ref{fig:compact_selection}, but are selected with an angular separation of $5\arcsec \leq r \leq 7\arcsec$. We select double sources of approximately equal lobe brightness in the orange box. See Figure \ref{fig:doubles_mosaic} for a representation of these sources appearance in FIRST. Also see Table \ref{tab:samples_used} for specific details on sample selection parameters for double sources.}
        \label{fig:doubles_selection}
    \end{figure}

    We correlate the flux ratio $\mathcal{R}$ with the compactness parameter $\mathcal{C}$ to select samples for stacking. An example of the sample selection process is shown in Figures \ref{fig:compact_selection} and \ref{fig:doubles_selection}. In both figures we show sources selected from a flux bin of $50.0 \leq S_{NVSS} \leq 70.0$ (flux bin 1, Table~\ref{tab:samples_used}). In Figure \ref{fig:compact_selection} we select single sources by only accepting sources if the FIRST position is within 3$\arcsec$ of the NVSS position ($0 \leq r \leq 3\arcsec$). As expected, most sources are distributed around a flux ratio of $\mathcal{R} = 1$. Single sources are further divided into 5 sub-samples with varying compactness $\mathcal{C}$ ranges which are represented by the dark blue, blue, cyan, green, and yellow boxes respectively. These samples are referred to as \textit{Single a} (most compact, all flux bins) to \textit{Single e} (most extended, all flux bins), where necessary accompanied by an index number for the flux bin (1 to 7, e.g. \textit{Single 3a}). We may also refer to all \textit{Single} samples in flux bin 1 as \textit{Single 1}. 
    
    Similarly, in Figure \ref{fig:doubles_selection} we select double sources by only accepting sources if the FIRST position is considerably different from the NVSS position, in this case between $5\arcsec$ and $7\arcsec$. Now most sources are distributed near $\mathcal{R} \approx 2$. This is because there is another FIRST source within the synthesised NVSS beam. A sample of double sources is then selected in the orange box with flux ratio $1.7 \leq \mathcal{R} \leq 2.3$ and compactness $0.9 \leq \mathcal{C} \leq 1.4$. We call this the \textit{Double a} sample (Table~\ref{tab:samples_used}). Sources to the left and right of $\mathcal{R} \approx 2$ (orange box) are still likely double sources but asymmetrical in brightness. This process is then repeated for a position difference of $7\arcsec \leq r \leq 9\arcsec$. This is the \textit{Double b} sample. Where necessary the \textit{Double} samples are written with a flux bin number in analogy with the \textit{Single} samples.

    Using the method described in Figures \ref{fig:compact_selection} and \ref{fig:doubles_selection}, $49$ samples were selected across a NVSS flux range of $\sim6.6 \leq S_{NVSS} \leq 70.0$ mJy. Relevant information on the samples and their respective selection criteria is summarised in Table \ref{tab:samples_used}. The sample sizes range from several hundred to several thousand, and contain a total of 67,967 sources. Due to the nature of the selection algorithm, a small number of \textit{Double} sources were included in both angular separation selections. However, this only occurs for $\approx 1.6\%$ (62/3850) of all double sources selected for stacking and can be considered statistically negligible for our purposes.

\begin{figure}[h]
        \centering
        \includegraphics[trim={0.25cm 0.325cm 0.25cm 0.25cm},clip,width=\linewidth]{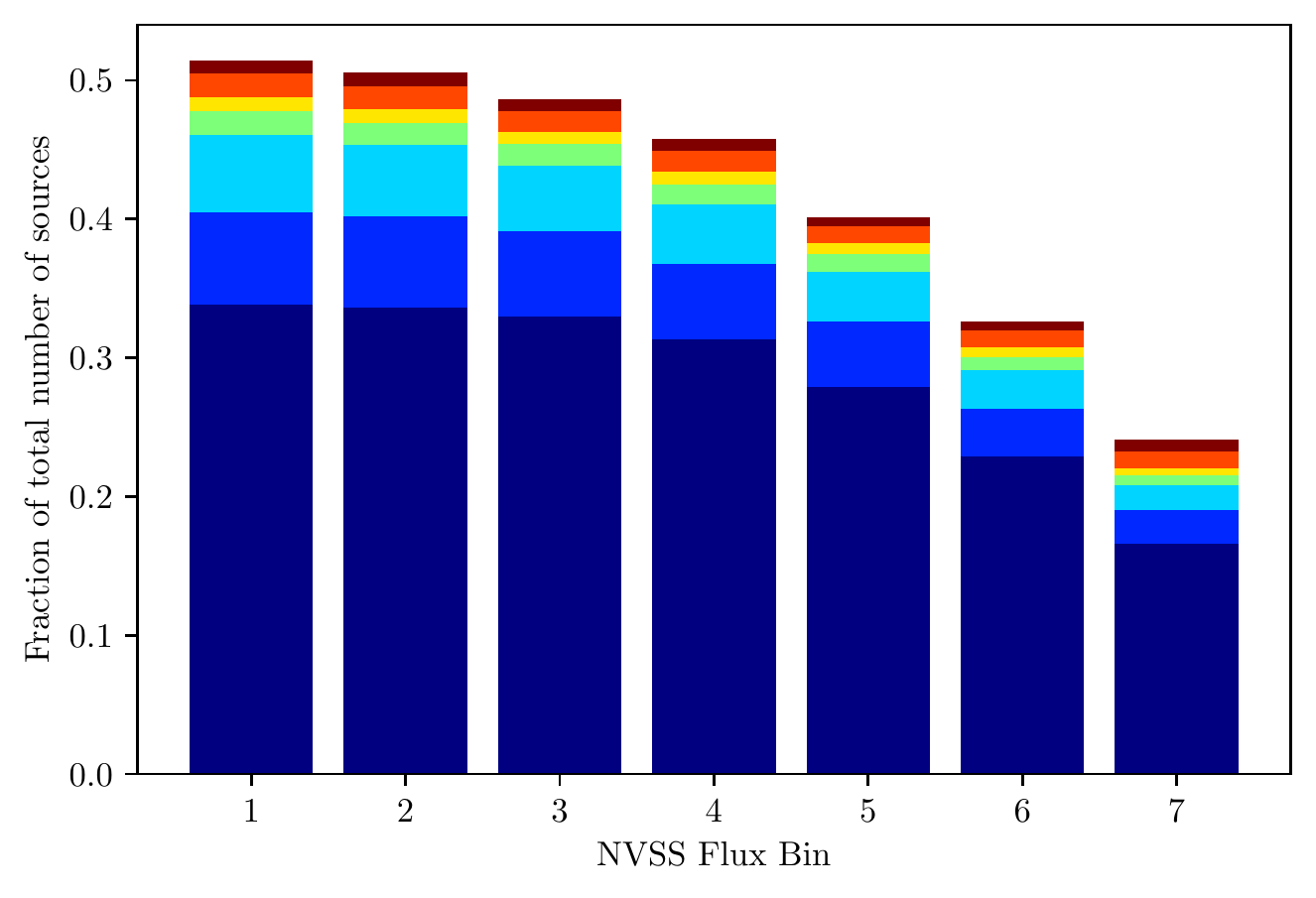}
        \caption{Fractions of the number of sources selected in each sample compared to the total number of sources in each flux bin. The histogram is stacked in order from most compact (bottom) to least compact (top), which corresponds to the \textit{Single a} (dark blue), \textit{b} (blue), \textit{c} (cyan), \textit{d} (green), and \textit{e} (yellow) samples as listed in Table \ref{tab:samples_used}. The \textit{Double a} and \textit{Double b} samples are at the very top in orange and red, respectively.}
        \label{fig:NVSS_souce_counts_bar_graph}
    \end{figure}

    Figure \ref{fig:NVSS_souce_counts_bar_graph} illustrates the fraction of all available sources that are actually included in our analysis, and the fraction of each sample. The figure compares the number of sources selected for stacking in each sample divided by the number of NVSS sources within the footprint of the FIRST survey for each flux density bin. In the brightest flux bin we obtain $\sim51.43$\% of all available sources, and $\sim24.09$\% of all sources in the faintest flux bin. We describe the morphology of the selected samples before we show a random selection of sources that were not included in any of our samples.
    
    The most compact sources ($0.9\leq\mathcal{C}\leq 1.15$), corresponding to the dark blue box of Figure \ref{fig:compact_selection}, represent the vast majority of the sources selected in each flux bin for stacking with sample sizes on the order of $\sim10^{3}$ and $\sim10^{4}$. It is also evident that the fraction of sources that we select in each bin becomes progressively smaller for fainter flux bins. This decrease may be due to sources being removed from the selection boxes due to noise or position errors. The efficacy of the sample selection limits the flux density range of this work, not our ability to detect a signal in stacked polarized intensity.
    
\begin{figure}[h]
        \includegraphics[trim={2.0cm 1.5cm 2.0cm 1.5cm},clip,width=\linewidth]{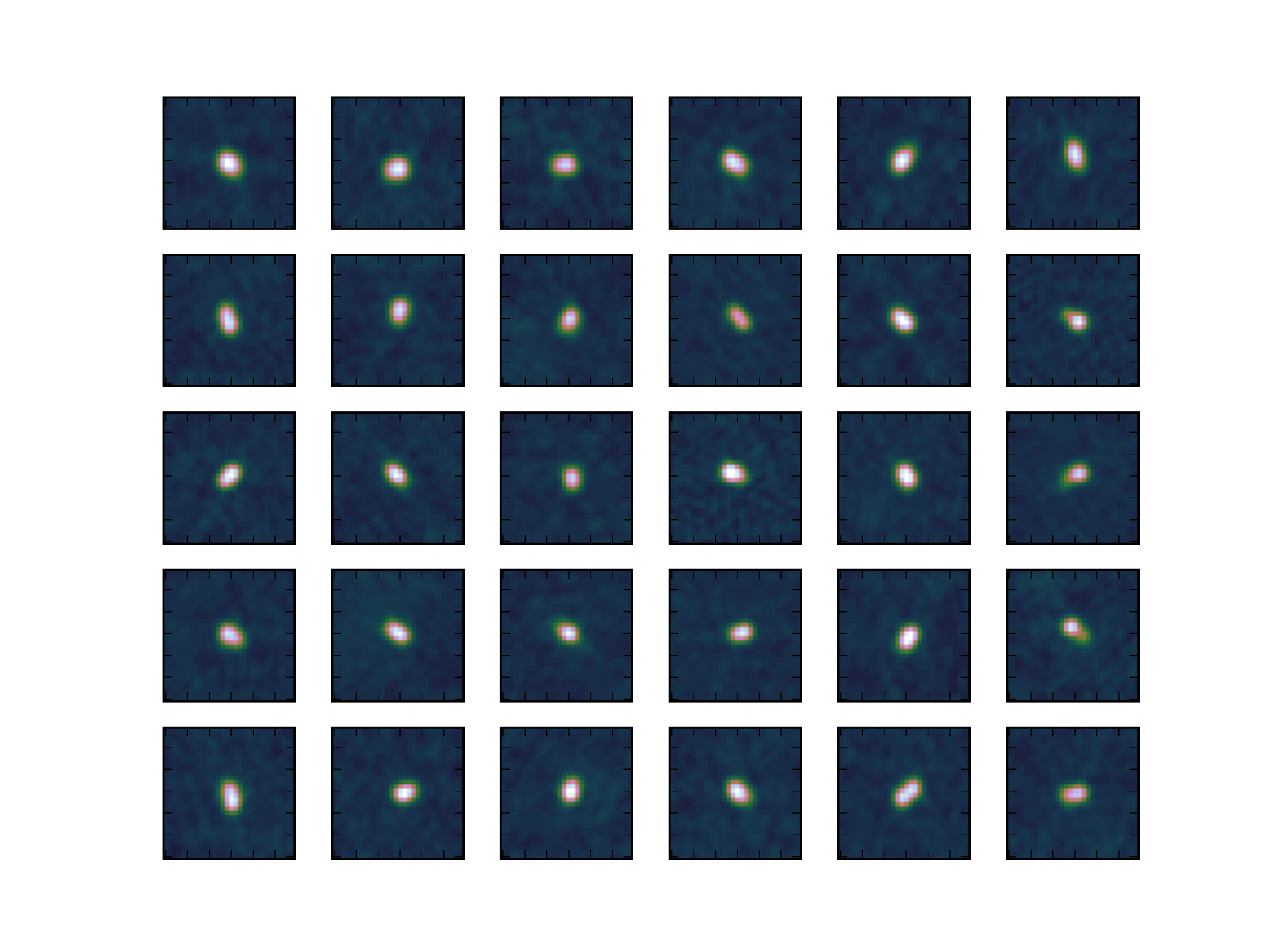}
        \caption{$\rm 5 \times 6$ mosaic of single sources from flux bin 5 ($13.0 \leq S_{1.4} \leq 18.2$ mJy) that are just resolved in FIRST, \textit{Sample 5b} (Table~\ref{tab:samples_used}, $1.15 \leq \mathcal{C} \leq 1.4$). Each individual image is a $54\arcsec \times 54\arcsec$ section of the sky centered on the position in the NVSS source catalog. }
        \label{fig:extended_mosaic_1}
    \end{figure}

\begin{figure}[h]
        \includegraphics[trim={2.0cm 1.5cm 2.0cm 1.5cm},clip,width=\linewidth]{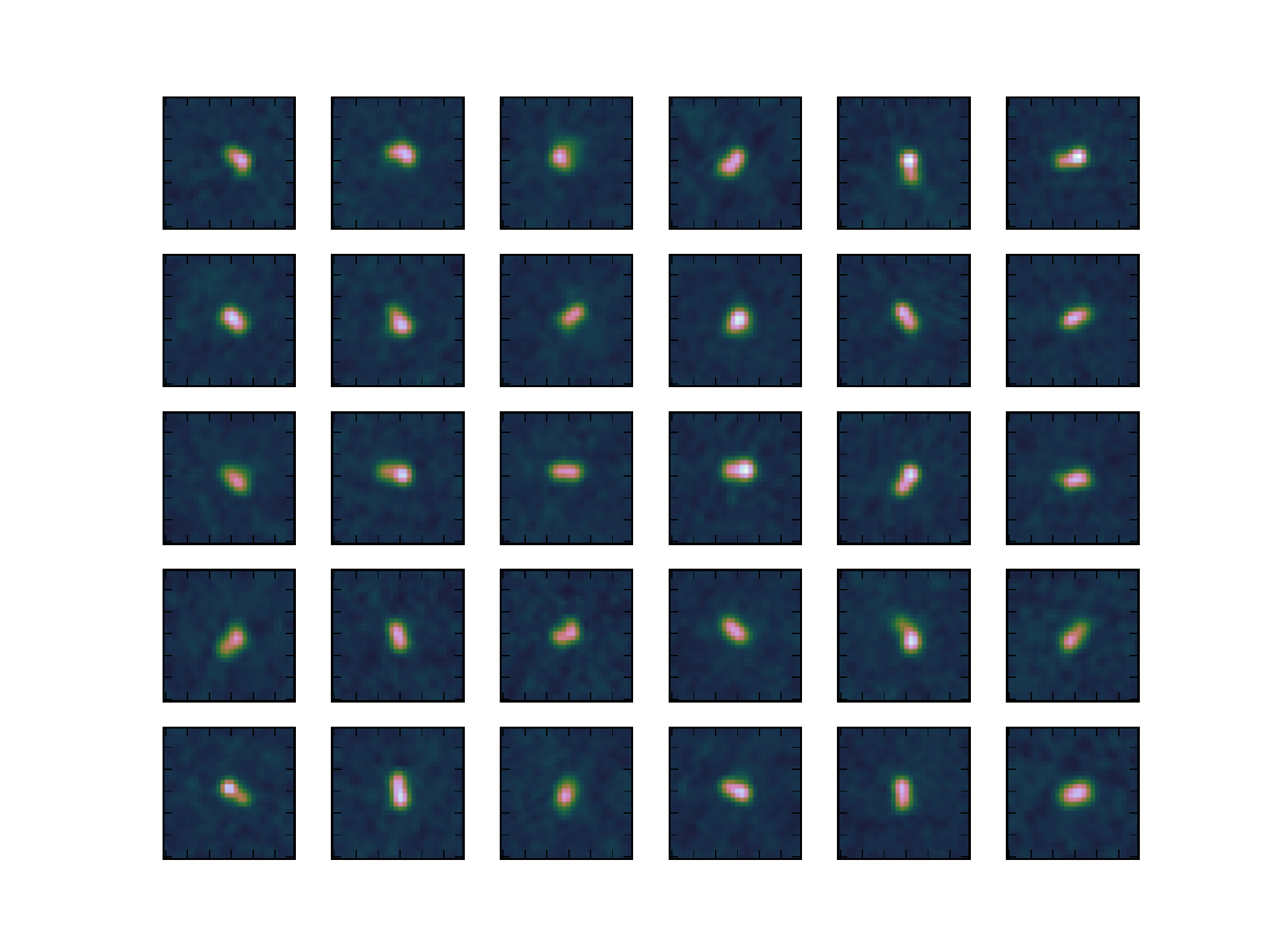}
        \caption{Mosaic of single sources that are well resolved in FIRST, \textit{Sample 5d} (Table~\ref{tab:samples_used}, $1.9 \leq \mathcal{C} \leq 2.4$, flux bin 5). The sources are shown with the same angular scale and intensity range as those in Figure~\ref{fig:extended_mosaic_1}.}
        \label{fig:extended_mosaic_2}
    \end{figure}

\begin{figure}[h]
        \includegraphics[trim={2.0cm 1.5cm 2.0cm 1.5cm},clip,width=\linewidth]{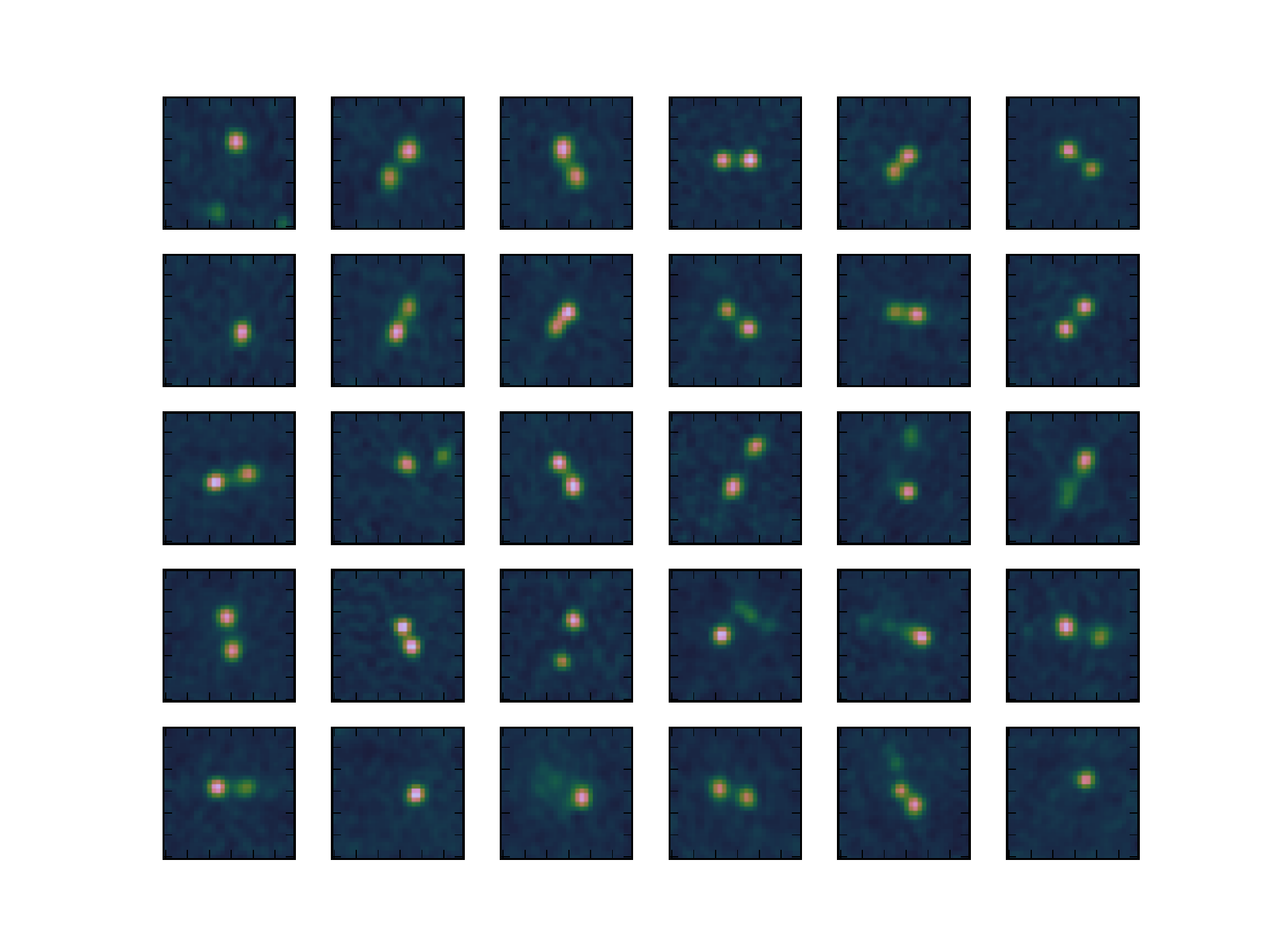}
        \caption{Mosaic of randomly selected double sources from sample \textit{Double 5a} (Table~\ref{tab:samples_used}, $5\arcsec \leq r \leq 7\arcsec$, flux bin 5). All other details are the same as those in Figures \ref{fig:extended_mosaic_1} and \ref{fig:extended_mosaic_2}.}
        \label{fig:doubles_mosaic}
    \end{figure}

    Examples of the appearance of these samples as seen in FIRST are provided in Figures \ref{fig:extended_mosaic_1}, \ref{fig:extended_mosaic_2}, and \ref{fig:doubles_mosaic}. Each figure is comprised of a $5 \times 6$ mosaic of $54\arcsec \times 54\arcsec$ images of sources selected from the $13.0 \leq S_{1.4} \leq 18.0$ mJy flux bin 5. This representation illustrates the composition of the samples in a way that any peculiar sub-class in the sample that represents $10\%$ or more of sources in the sample should be represented by a few examples in these figures.  The most compact samples examined in this paper are $\mathcal{C} \leq 1.15$, which are essentially unresolved in FIRST, trivially appearing as a point source. Sources that are just resolved in FIRST can be seen in Figure \ref{fig:extended_mosaic_1} for a compactness range of $1.15 \leq \mathcal{C} \leq 1.4$ (\textit{Single 5b}). Sources that are even more resolved ($1.9 \leq \mathcal{C} \leq 2.4$; \textit{Single 5d}) are shown in Figure \ref{fig:extended_mosaic_2}. With this visual inspection we verify that our ranges of $\mathcal{C}$ effectively sort sources by angular size, with no significant dilution of the samples by noise. These figures also suggest that none of our samples is close to the confusion limit.
    
    For comparison, Figure~\ref{fig:doubles_mosaic} shows a mosaic of the double radio sources (\textit{Double 5a}). As expected, each image in the mosaic consists of two reasonably symmetric radio lobes near each other at various orientations. However, some sources such as those in panels (row, column) (2,1) and (5,2) appear to have only one lobe present or have very faint lobes. This suggests there is extended emission in some sources that is resolved out in FIRST, which is one of the challenges higher resolution surveys face as pointed out by \cite{Rudnick_Owen_2014}.

\begin{figure}[h]
        \includegraphics[trim={2.0cm 1.5cm 2.0cm 1.5cm},clip,width=\linewidth]{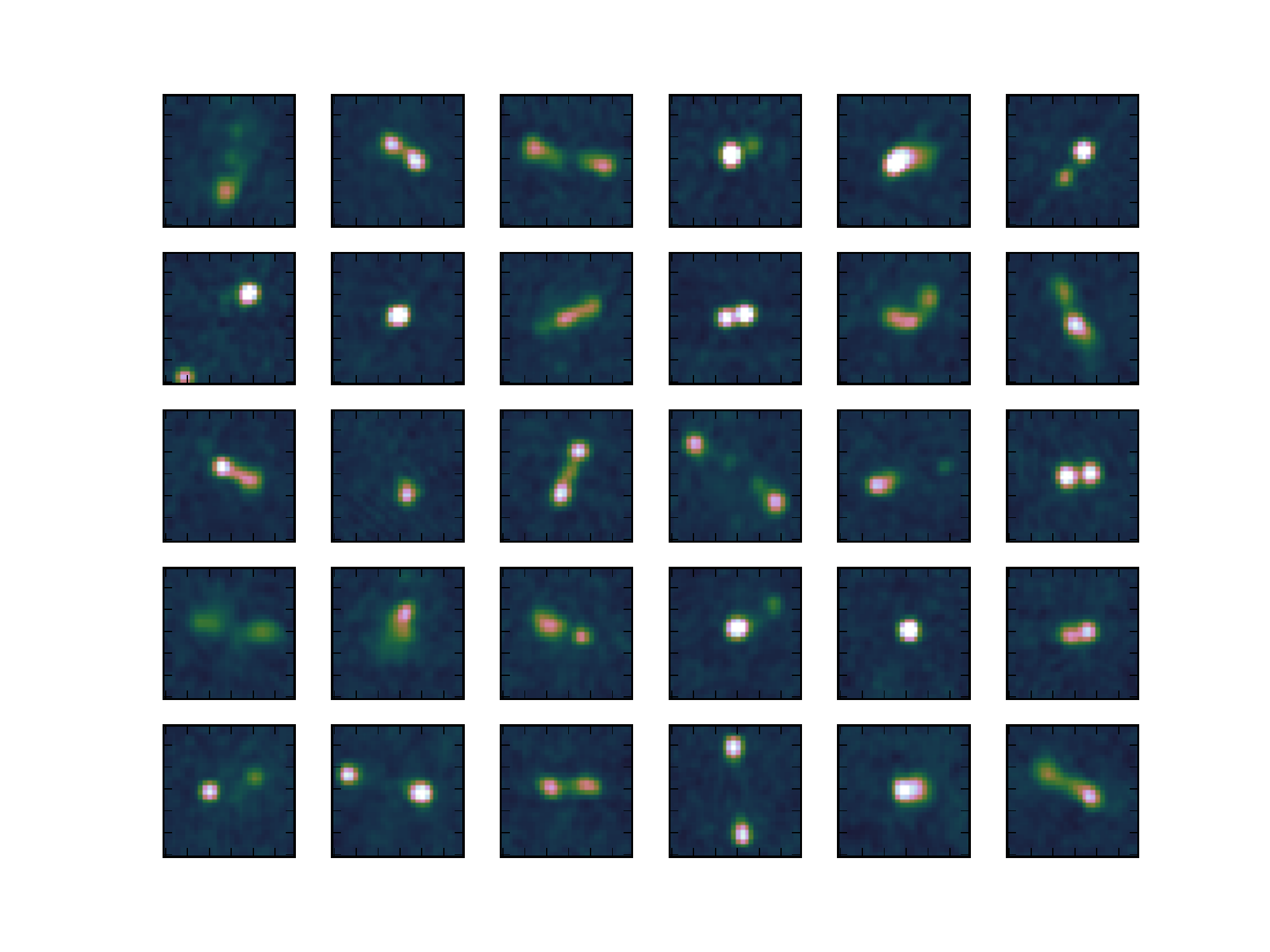}
        \caption{A random selection of 30 sources rejected by our sample selection for flux bin 3 (Table~\ref{tab:samples_used}). The panels are $54\arcsec$ on a side, illustrating that many of rejected sources are more extended than any in our samples. Smaller subsets of rejected sources consist of asymmetric doubles and compact sources whose FIRST flux density differs from the NVSS flux density, presumably because of variability. Other possible reasons for rejection are excessive flux errors related to image artifacts and occasional large position errors. At these flux density levels, the chance of a random alignment within the NVSS beam is small because the confusion limit is well below the brightness of the faintest samples.}
        \label{fig:unused_sources_mosaic}
    \end{figure}

    Figure~\ref{fig:unused_sources_mosaic} shows a random subset of the sources in flux bin 3 that were not selected. The panels are $54\arcsec$ on a side to allow comparison with the samples represented in Figures~\ref{fig:extended_mosaic_1}, \ref{fig:extended_mosaic_2}, and \ref{fig:doubles_mosaic}. About half of the rejected sources have larger angular size and more complex morphology than sources in our samples. On rare occasions, we find a source that is just one lobe of a much more extended source. This may be the case for the first and the fifth source in the top row of Figure~\ref{fig:unused_sources_mosaic}. The larger angular size of these sources would complicate interpretation of the stacked intensities. Our sample selection also rejects asymmetric double sources. These are interesting in their own right, but stacking of asymmetric double sources is a significant separate investigation. We retain the symmetric {\it Double} samples because their components have the same $\mathcal{C}$ as our most compact \textit{Single} samples.

    We also see some rejected compact sources, most likely due to variability, but this is not a large fraction of the rejected sources. Rejection because of variability must be rare because we find few sources with $\mathcal{R} << 1$ beyond what can be attributed to noise (Figure~\ref{fig:compact_selection}). \cite{TGSS} find that $\mathcal{R}$ can decrease to $\sim 0.8$ for the faintest detectable sources in their survey. $\mathcal{R}<1$ occurs also for brighter sources, but asymptotically $\mathcal{R}$ approaches 1. We do not use sources close to the detection threshold of FIRST and NVSS, so we apply a lower limit $\mathcal{R} > 0.9$. The requirement of a significant position offset between FIRST and NVSS guards against contamination of the \textit{Double} samples by variable \textit{Single} sources.  By far the largest subset of rejected sources are extended radio galaxies with angular size comparable to or larger than the NVSS beam. This adds confidence that our \textit{Single} samples represent well-defined and reasonably complete sub-sets of the compact mJy radio source population.
    
    The representative angular size of sources in our samples may be derived using the new quick-look images of the VLA Sky Survey (VLASS), nominal frequency 3 GHz, angular resolution $2.5\arcsec$. The VLASS quick-look images are slightly under-sampled and have only limited application because of their preliminary calibration and imaging. The frequency is also higher, which may cause subtle effects imposed by differences in spectral index. However, all our samples are within the angular size range detectable by VLASS, and the $2.5\arcsec$ angular resolution is high enough to place meaningful limits on source size.

    Table~\ref{tab:results} lists $\theta_{1/2}$, the FWHM of the emission of the single-source samples derived from Gaussian fits to the mean-stacked VLASS images. For the largest two samples, the brightness distribution is not quite Gaussian and the images are less smooth because of the smaller sample size and the larger solid angle. The estimated error for the compact samples is about $0.1\arcsec$, while for the two most extended samples the error is approximately $1\arcsec$. The fitted FWHM is larger for the fainter samples compared to brighter samples with the same compactness $\mathcal{C}$, in a way that is consistent with NVSS position errors increasing from about $1\arcsec$ (FWHM) to $3\arcsec$ (FWHM) over the flux density range $70\ \rm mJy$ to $7\ \rm mJy$. 
    
    Deconvolved angular sizes were derived from the mean-stacked VLASS images by subtracting in quadrature a $2.5\arcsec$ (FWHM) circular Gaussian beam and $1\arcsec$ (FWHM) position errors for the brightest samples, with no further correction for under-sampling of the VLASS synthesized beam. This results in deconvolved FWHM sizes for the samples of \textit{Single 1} sources (50 mJy to 70 mJy) as $1.8\arcsec \pm 0.1 \arcsec$ ($0.9 \le \mathcal{C} \le 1.15$), $2.9\arcsec \pm 0.1 \arcsec$ ($1.15 \le \mathcal{C} \le 1.4$), $4.9\arcsec \pm 0.2\arcsec$ ($1.4 \le \mathcal{C} \le 1.9$), $6.5\arcsec \pm 1\arcsec$ ($1.9 \le \mathcal{C} \le 2.4$), and $8.2\arcsec \pm 1\arcsec$ ($2.4 \le \mathcal{C} \le 5.0$). These correspond to projected physical sizes of 15, 24, 41, 54, and 68 kpc, respectively for an angular size distance $1.72$ Gpc at $z=1$. We note that the median angular size at 70 mJy according to the relation of \citet{Windhorst_2003} is $7\arcsec$, which is large compared to the angular sizes found for our samples. This is a result of our sample selection, which is more targeted towards compact, single sources in the FIRST survey. We will return to this in Section~\ref{sec:discussion}.
    
\begin{figure}[h!]
        \centering
        \includegraphics[width=1.05\linewidth]{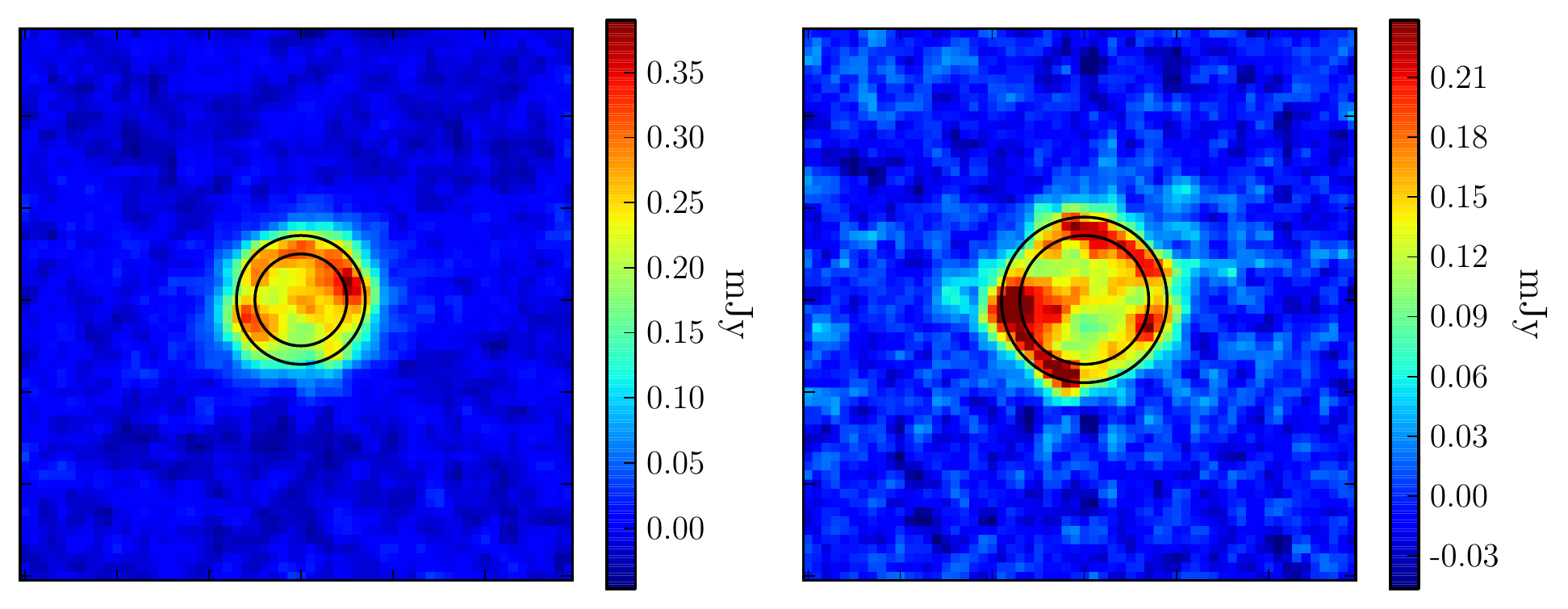}
        \caption{Mean stacked Stokes I (VLASS) images of the Doubles 5a (left) and 5b samples (right). Black circles are plotted with radius $5\arcsec$ and $7\arcsec$ for the \textit{Double 5a} sample and radius $7\arcsec$ and $9\arcsec$ for the \textit{Double 5b} sample. Both images are $60\arcsec \times 60\arcsec$ in dimension, which is the FWHM beam size of the images used in the stacking analysis.}
        \label{fig:VLASS_doubles}
    \end{figure}

    Mean-stacked VLASS images of the \textit{Double 5} samples are shown in Figure~\ref{fig:VLASS_doubles}. The ring morphology is the signature of a set of randomly oriented double sources with separation according to expectations from the sample selection. We find little or no evidence for a central core as a third source component. The symmetric appearance in individual VLASS and FIRST images suggests that the sample is dominated by edge-brightened sources with a FR II morphology. Only a few of our \textit{Double} sources appear as a compact core with a more extended jet or lobe. The mean stacked VLASS images show significant brightness within the annulus defined in the sample selection. We adopt an angular size of $12\arcsec$ (100 kpc at $z=1$) for the \textit{Double a} sample and an angular size of $16\arcsec$ (133 kpc) to the \textit{Double b} sample. In this paper, we are mainly concerned with the angular size of the \textit{Single} samples, which is derived in a different way. We only make qualitative reference to the relative sizes of the \textit{Single} and \textit{Double} samples. This is possible because our sample selection strictly requires one compact component in the FIRST catalog to be within the annulus, with a flux density that is approximately half the NVSS flux density, and no further requirement for the distribution of the remaining flux. There is no requirement whether the brighter or the fainter component matches the offset criterion with the NVSS position. Inspection of the sample (Figure~\ref{fig:doubles_mosaic}) shows that this results mostly in sources for which most of the flux density is in {\it two} compact components. Individual VLASS images suggest a FR II morphology for most sources in our samples of doubles, so we tentatively interpret the two components as hot spots that trace the extremities of the source. The source list was filtered to avoid the same source appearing twice in the target list.
    
    Our samples of \textit{Doubles} were selected for comparison with the \textit{Single} source samples as a function of angular size. The sample selection targets \textit{Doubles} with at least one compact component. Figure~\ref{fig:doubles_selection} suggests that there are other double sources with a more unequal flux density ratio of the components, and perhaps more diffuse components. This is significant because our \textit{Double} sample selection may favour sources with jet axis in the plane of the sky whose opposite sides are the least subject to the Laing-Garrington effect \citep{Laing_1988,Garrington_1988}. The \textit{Double} samples are by selection almost exclusively sources with FR II morphology, while the \textit{Single} samples probably include sources with different morphology that remains undetermined with the available data. This should be taken into account when comparing the \textit{Double} samples with the \textit{Single} samples. We will return to this topic in the discussion. 
    
    \begin{figure}[h]
        \includegraphics[width=\linewidth]{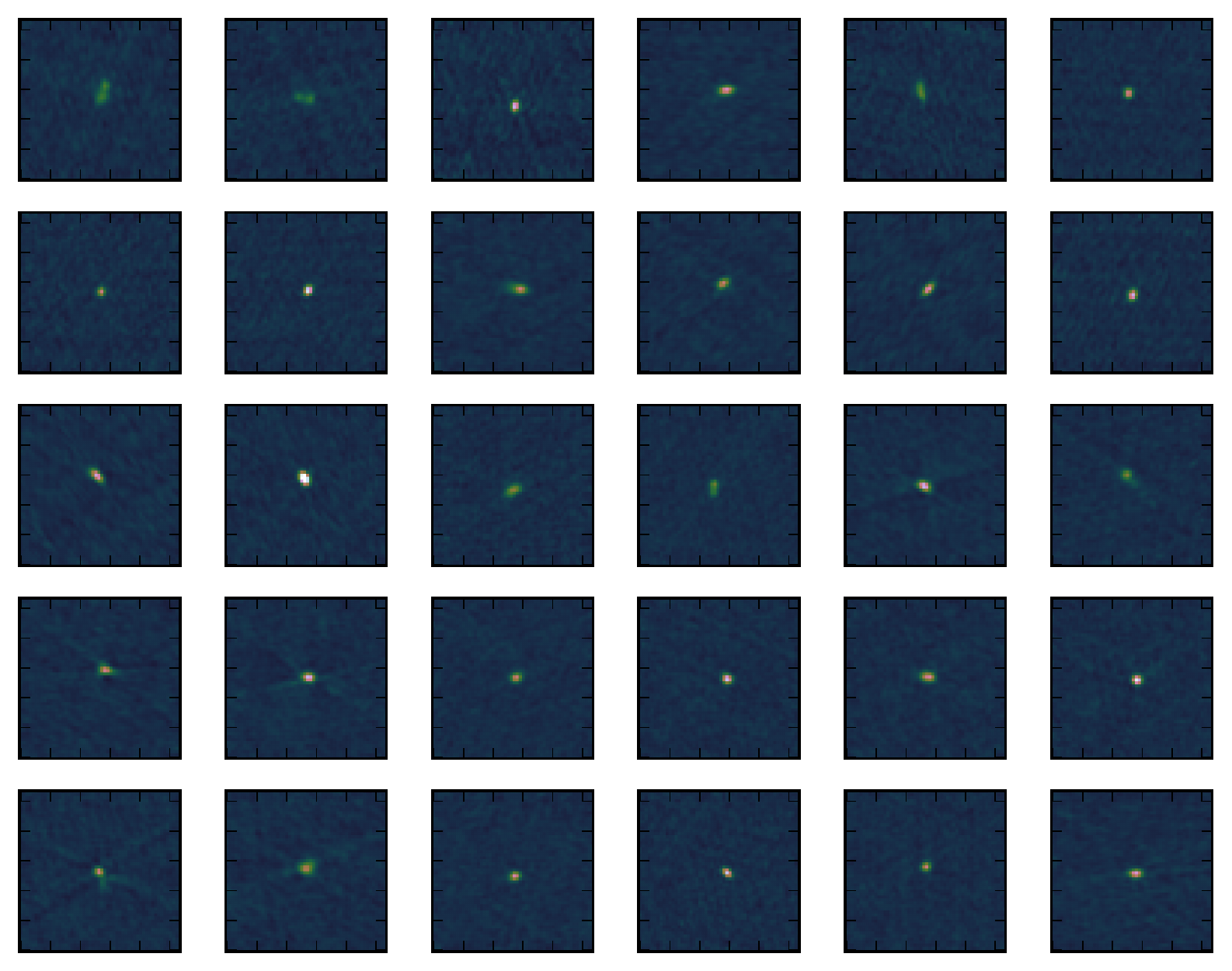}
        \caption{Mosaic of VLASS images consisting of randomly selected compact sources from the \textit{Single 5a} sample ($0.9 \leq \mathcal{C} \leq 1.15$). Each image is a $54\arcsec \times 54\arcsec$ section of the sky centered on the position listed in the NVSS source catalog. The intensity scale was chosen to highlight differences between individual sources.}
        \label{fig:vlass_mosaic_compact}
    \end{figure}
    
    Unlike the \textit{Double} samples, the mean-stacked VLASS images of the \textit{Single} samples are clearly brightest at the center, defined by the NVSS position. Inspection of the VLASS images of individual sources shown in Figure~\ref{fig:vlass_mosaic_compact} shows a linear morphology that is often, but not always edge-darkened, reminiscent of a poorly resolved source with FR I morphology. Blended lobes of an FR II source may give the same appearance. Approximately half of the \textit{Single a} sources in Figure~\ref{fig:vlass_mosaic_compact} shows asymmetry with a brighter side and a fainter side, and half is quite symmetric, elongated in one direction and unresolved in the orthogonal direction. The \textit{Single a} sample also contains sources that appear compact double sources at the resolution of VLASS, so by no means do we suggest that this is a homogeneous morphological class. Very few sources in our samples appear to be resolved in two dimensions in VLASS images.
    
    It is beyond the scope of this paper to pursue optical or near-infrared identifications of these samples. Inspection of SDSS DR7 images made available through the \textit{Skyview} virtual observatory \citep{skyview_1998} of a few randomly selected objects from each sample indicates that for all samples the host galaxies must be near or beyond the photometric limit of the SDSS ($r$ magnitude $\approx 22$. This corresponds to absolute magnitude $M_r = -20.3$ at $z = 0.5$ and $M_r = -22.2$ at $z = 1$. \citet{Bornancini_2010} presented a detailed study of radio galaxies in the SDSS in relation to spectral index and environment. Considering that host galaxies of radio loud AGN tend to be luminous elliptical galaxies, we get a crude indication that all our samples are dominated by radio sources at significant redshift. Most importantly, it is clear that our most compact sample does not include a large number of low-redshift star forming galaxies.

\section{Results} \label{sec:results}
    \subsection{Derivation of the true sample median $\Pi_{0, \rm med}$}
    \label{sec:results-stat}
    The main objective of this paper is to further investigate the results of \cite{Rudnick_Owen_2014}. In this paper we follow the procedure for stacking total and polarized intensity as outlined by \cite{stil_keller_george_taylor_2014}, which is performed using the Python Astronomical Stacking Tool Array (PASTA) by \cite{PASTA}. Table \ref{tab:results} lists the median $S_{\rm NVSS}$ flux densities for each stacked image, the number of sources included in each stack ($N_{\rm stack}$), and the results of subsequent analysis of the stacking. 
    
\begin{figure}[h]
        \centering
        \includegraphics[width=0.65\linewidth]{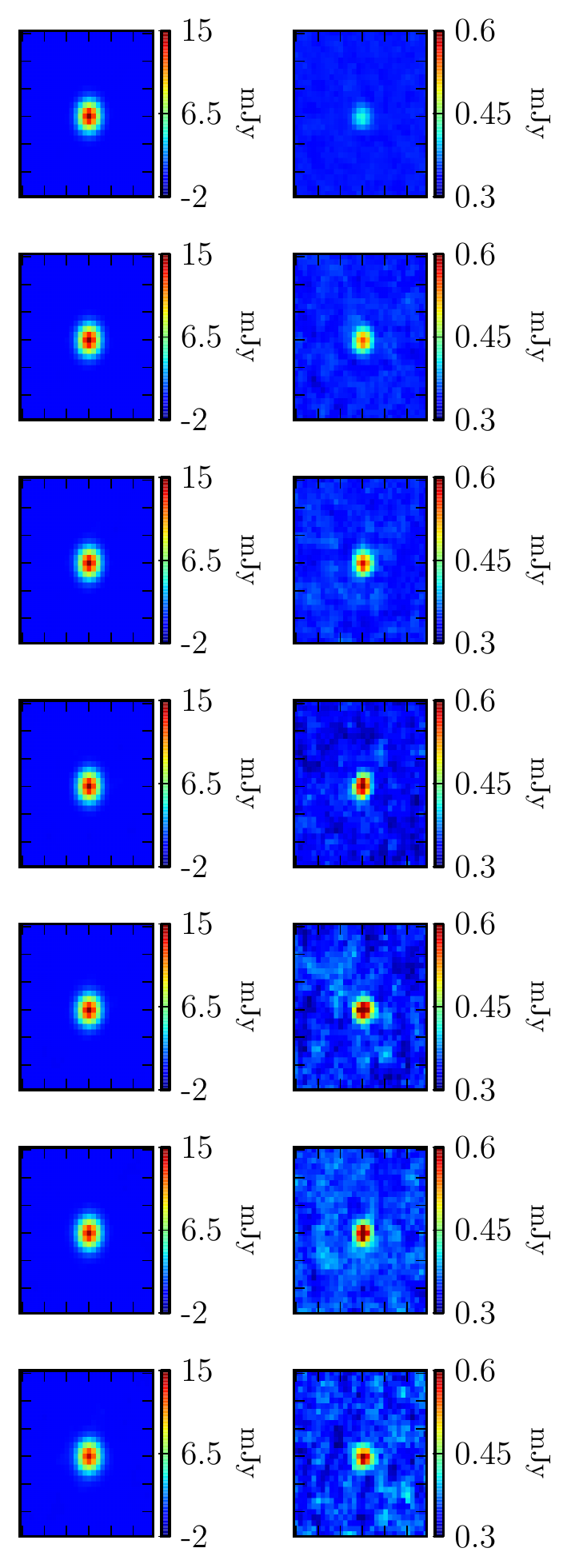} 
        \caption{Stacked images (NVSS) of total intensity (left column) and polarized intensity (right column) for the \textit{Singles 5a} (top row) to \textit{Singles 5d} samples and the \textit{Double 5a} and \textit{Double 5b} (bottom row) samples. Angular size of the source increases from top to bottom. The differences between the stacked Stokes I images are too subtle to be seen in this representation. The stacked polarized intensity images display a significant increase in the median polarized signal with angular size of the source, most notably for the more compact sources. The noise in stacked polarized intensity is higher for smaller samples, resulting in a lower signal to noise ratio for samples with stronger median polarized signal in this work.}
        \label{fig:sample_PI_vs_I_images}
    \end{figure}

    Figure \ref{fig:sample_PI_vs_I_images} shows raw stacked images in total (left column) and polarized (right column) intensity from flux bin 5 with a median flux density $S_{\rm NVSS} = 15.35$ mJy. Each row of images is a sub-sample which corresponds to a different compactness $\mathcal{C}$ selection. Source angular size increases from top to bottom, the last 2 rows being the \textit{Double} samples. The median peak total intensity of the images shown is $14.99 \pm 0.05$ mJy and varies between each sub-sample by up to a maximum $1.59\%$ difference from the median value (See Table \ref{tab:results}). By comparison, the Stokes I background noise varies between $\sim0.02$ and $\sim 0.07$ mJy beam$^{-1}$. The noise in Stokes I stacked images is raised by residual side lobes of emission below the clean threshold of the NVSS. The most compact sample for the flux bin shown in Figure \ref{fig:sample_PI_vs_I_images} has $8202$ sources, while the most extended sample has only $223$ sources. 
    
    While the total intensity images in Figure \ref{fig:sample_PI_vs_I_images} do not vary significantly, there are clear visible differences in the median stacked polarized intensity images. Notice the much fainter stacked polarized median intensity of $p_{\rm med}=0.41$ mJy for the sources unresolved in FIRST at $5\arcsec$. There is a considerable jump in polarized intensity to $p_{\rm med}=0.55$ mJy as soon as a source becomes resolved in FIRST, with subtle visible increases in strength as a source becomes more extended. Additionally, we can see that the double sources in the last two rows have the strongest polarized intensity signal at $p_{\rm med} = 0.64$ mJy. We find similar results in the raw stacked images across all flux bins.  
    
    The mean background level in stacked polarized intensity ($p_{\rm med}$) is always positive as a result of the polarization bias in the absence of signal. The average noise in the Stokes Q and U images from the NVSS is 0.29 mJy beam$^{-1}$ \citep{NVSS}. In each of the stacked polarized intensity images, the median off-source background is approximately 0.365 mJy beam$^{-1}$, which is close to the expected median of the Rayleigh distribution $\sigma_{QU} \sqrt{2 \ln 2} = 0.341$ mJy beam$^{-1}$. In the present analysis, we use local noise values, derived from the actual survey images, for each source.

    \begin{figure}[h!]
        \includegraphics[trim={0.25cm 0.15cm 1.5cm 1.25cm},clip,width=\linewidth]{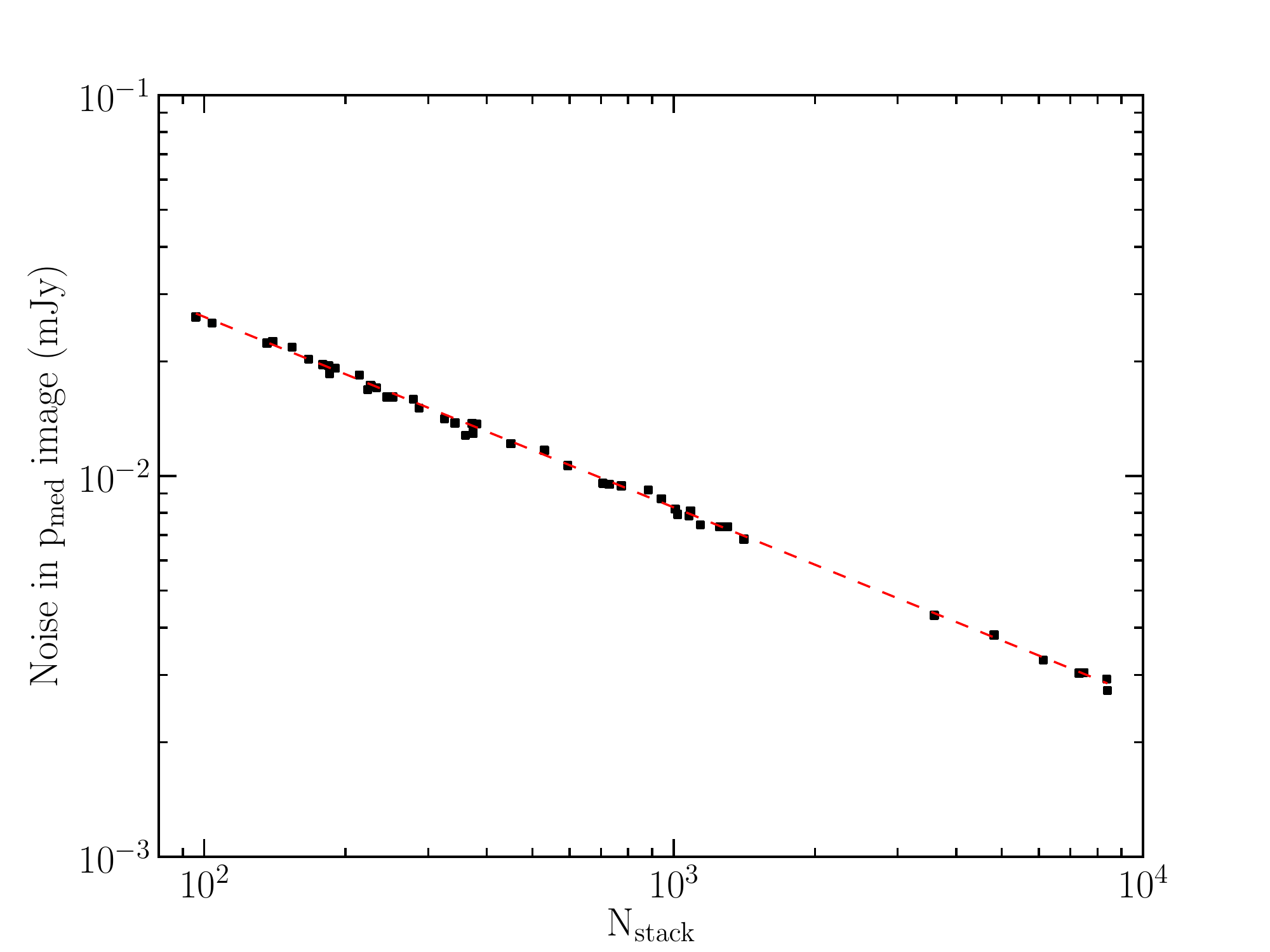}
        \caption{Relationship between the noise in the $p_{\rm med}$ stacked images and the number of sources used in the stacked sample ($N_{\rm stack}$). The data points represent the standard deviation of the background noise of each stacked image. The dashed red line is a fit of the data where $\rm{noise} = B \times N^{\beta}$ with $B = 0.260 \pm 0.005$, and $\beta = -0.500 \pm 0.003$, for a domain of $96 \leq N_{\rm stack} \leq 8397$.}
        \label{fig:noise_sample_size}
    \end{figure}
    
    The noise level of the stacked polarized intensity images as a function of sample size is examined by offsetting the declination values from the original target sources in each sample by $1\degr$. These new offset target lists are then stacked again, effectively creating median stacked images of the background noise surrounding the area of the source. Figure~\ref{fig:noise_sample_size} visualizes the off-source background noise as a function of the stacked sample size ($N_{\rm stack}$), which can be fitted by a power law of the form $\mathrm{noise} = B N^{\beta}$, where $B = 0.260 \pm 0.005$, and $\beta = -0.500 \pm 0.003$. This fit is represented by the dashed red line in the figure. 

    All of the sources in each of the selected samples are distributed throughout the survey. This means that the differences in the raw median stacked images such as those shown in Figure \ref{fig:sample_PI_vs_I_images} are not simply the result of an artifact of noise or other data bias, instead being the result of real astrophysical differences. Therefore, it can already be concluded in a qualitative way that the percent polarization of extragalactic sources is strongly related to their structure, with extended sources on a scale of $\gtrsim 5\arcsec$ having a higher median percent polarization than more compact sources.
    
    \subsection{Monte Carlo simulations}
    The noise bias and statistical errors related to polarization are modeled with Monte Carlo simulations of the stacking as described in Section \ref{subsec:stacking}. The true median fractional polarization $\Pi_{0,\rm med}$ for each sample derived from the Monte Carlo simulations is listed in Table~\ref{tab:results} with errors derived from the variance in the Monte Carlo simulations. The errors are larger for the \textit{Single b - e} samples and the \textit{Double} samples due to the smaller sample sizes. We find very significant differences in median fractional polarization in relation to structure. Before we discuss these in more detail, we must return to the effect of the assumed distribution function of $\Pi_0$ on the derived sample median $\Pi_{0, \rm med}$.
    
    \begin{figure*}[t!]
        \centering
        \includegraphics[width=0.7\linewidth]{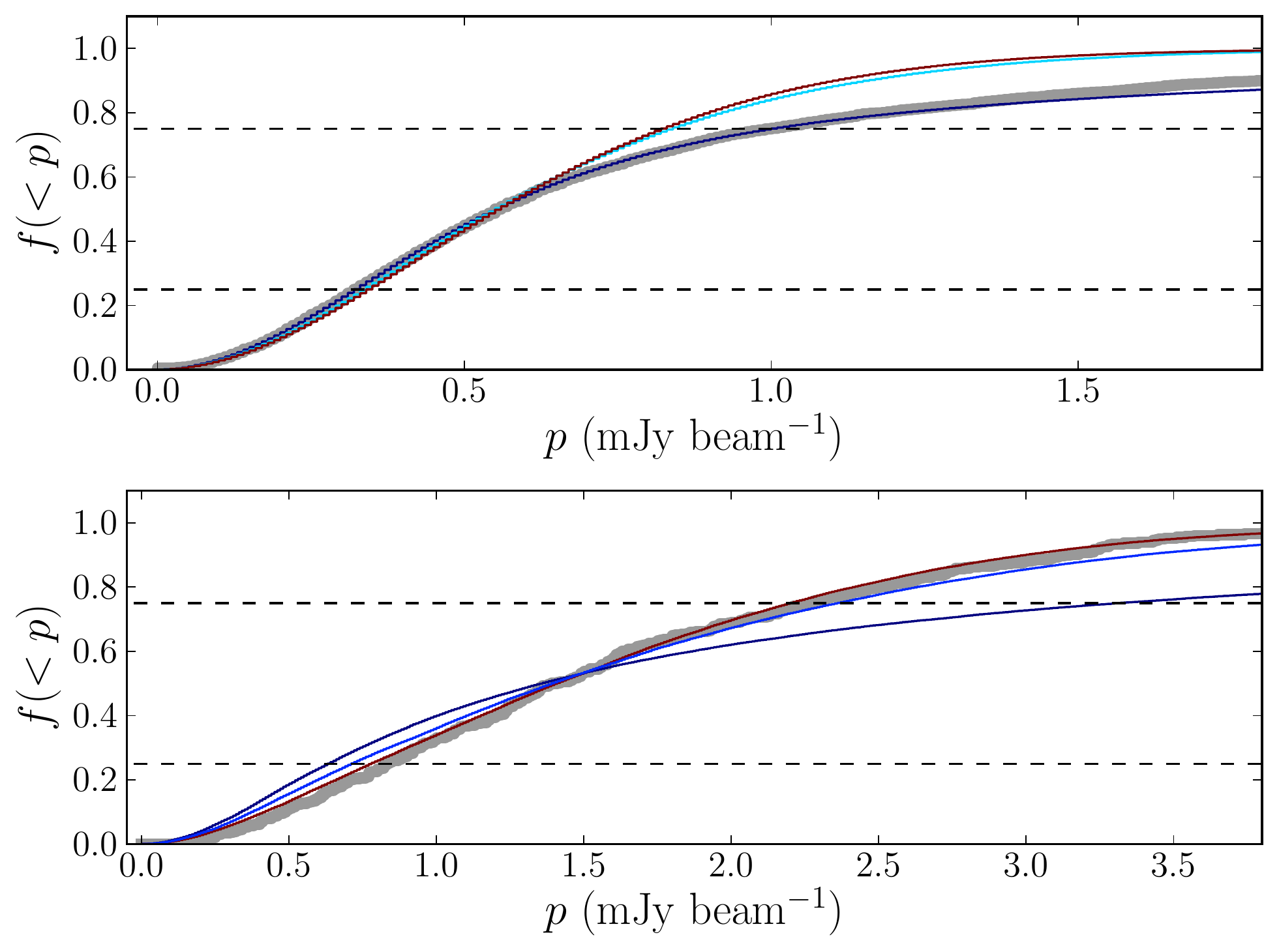}
        \caption{Cumulative distributions of observed polarized intensity $p$ and model distributions for the \textit{Single 2a} sample (top) and the combined \textit{Double 2} sample (bottom). The data are represented by the thick gray curves. The model curves are colored according to the sample for which the adopted distribution provides a good fit, following the color scheme adopted in other figures in this paper. The dark blue curves assume the $\Pi_0$ distribution that applies to the \textit{Single a} sample (Equation~\ref{eq:compact_source_dist}). The blue and cyan curves assume a Gaussian distribution, appropriate for the \textit{Single b} (blue) and \textit{Single c} (cyan) samples. The red curve assumes the $\Pi_0$ distribution in Equation~\ref{eq:double_source_dist} that applies to the \textit{Doubles} sample. Model distributions differ only in the adopted intrinsic distribution of fractional polarization $\Pi_0$. The median of all distributions was scaled such that the modeled median $p_{\rm med}$ reproduces the median of the data. Since the blue and cyan curves would be the same, only one is shown in each panel. The dashed horizontal lines mark the $25\%$ and $75\%$ quartiles of the distributions. This figure illustrates how the ratio of the quartiles $p_{75\%}/p_{25\%}$ is used to verify that the model distributions agree with the data in a manner that is robust against outliers that populate the tail of the distribution.}
        \label{fig:p_cumdist_2panel}
    \end{figure*}
    
    \begin{figure*}[t!]
        \centering
        \includegraphics[width=0.95\linewidth]{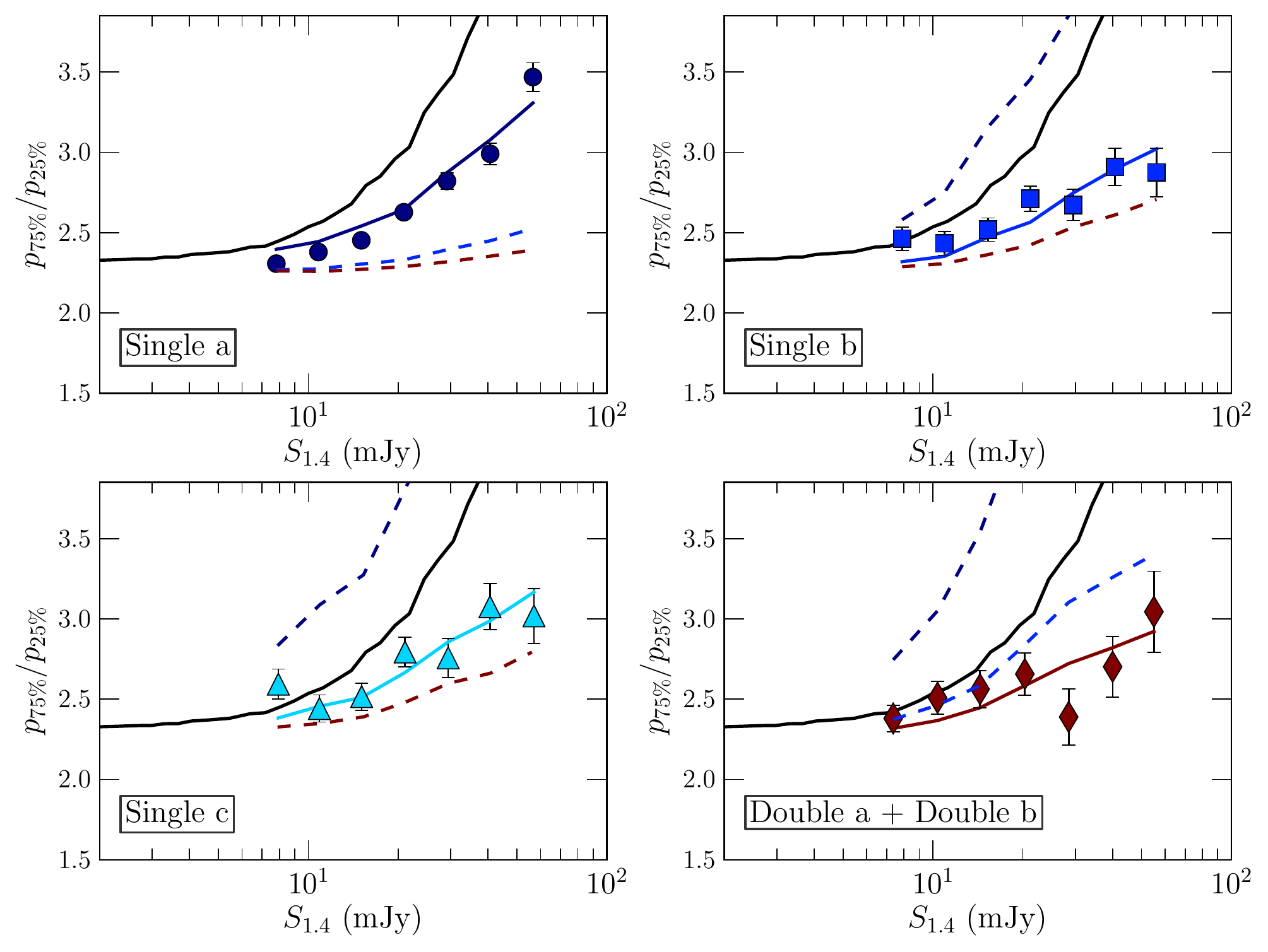}
        \caption{Quartile ratios of polarized intensity for the samples \textit{ Single a} (top left), \textit{Single b} (top right), \textit{Single c} (bottom left), and the combined sample \textit{Double a}$+$\textit{Double {b}} (bottom right). Filled symbols indicate the quartile ratios of the data. Solid curves in the color of the symbols indicate the quartile ratios of the Monte Carlo simulations of the stacks. The error bars indicate the standard deviation of the quartile ratio for a single stack, derived from these simulations. The significance of differences in the $\Pi_0$ distribution is illustrated by the dashed curves. These show the quartile ratios from a Monte Carlo analysis using the $\Pi_0$ distribution function of one of the other samples, indicated by the color of the curve. The colors indicate the same samples as the colors in Figure~\ref{fig:NVSS_souce_counts_bar_graph}. The blue dashed curve in the upper left panel represents the quartile ratios from a Monte Carlo analysis of the \textit{Single a} stacks, assuming the $\Pi_0$ distribution of the \textit{Single b} sample.  The dark red dashed curve in the upper left panel shows the Monte Carlo analysis of the \textit{Single a} stacks, assuming the $\Pi_0$ distribution of the \textit{Double} sample. The dashed curves in other panels are corresponding analogues. The $\Pi_0$ distribution functions of the \textit{Single b} and \textit{Single c} are both Gaussians, so only one is shown. Deviation of the dashed curves from the data in the same panel indicates that, while the observed $\Pi_{0,\rm  med}$ is reproduced, the remainder of the cumulative distribution of polarized intensity is not well reproduced. The black curve in each panel is the quartile ratio curve for the entire NVSS sample, as shown in Figure 9 of \cite{stil_keller_george_taylor_2014}.}
        \label{fig:quartile_ratio-fig}
    \end{figure*}
    
    \citet{stil_keller_george_taylor_2014} pointed out that samples with the same true median polarized intensity will have somewhat different stacked median polarized intensity depending on the true distribution of polarized intensity of the sample. This happens because of the non-linear relation between the expectation value of polarized intensity and the true signal at low signal to noise ratio. The distribution on the derived $\Pi_{0,\rm med}$ introduces a systematic effect that is typically several times the statistical noise in the stacked polarized intensity, but it is only a fraction of the polarization bias correction itself \citep[c.f.][Figures 4 and 6]{stil_keller_george_taylor_2014}. A distribution of true polarized intensities implies that the polarized signal to noise ratio is different for every source, even if the noise is perfectly uniform in the survey. The non-linear relation between observed and true polarized intensity yields a disproportionately higher polarized intensity for the sources in the sample that have a higher $p_0$. Unless the signal to noise ratio is high, the true distribution of $p_0$ skews the distribution of observed $p$. For sufficiently large samples, this effect can be detected and exploited to derive or approximate the distribution of $p_0$ \citep{stil_keller_george_taylor_2014}.
    
    Figure~\ref{fig:p_cumdist_2panel} illustrates the effect of the distribution of intrinsic $p_{0,i}$ on the distribution of observed $p_i$ for two of our samples. The thick gray curves represent the cumulative distributions of observed polarized intensity $p$ for the \textit{Single 2a} sample (top) and the combined \textit{Double 2} sample (bottom). We do Monte Carlo realizations of the stack that assume a distribution function of fractional polarization that is scaled to vary $\Pi_{0,\rm med}$ as a fitting parameter. Model distributions of $p_i$ are then generated for the specific sample, by reading the Stokes I intensity and local noise level for each source. We apply a fractional polarization drawn from the assumed distribution, to be multiplied by the Stokes I intensity of a real source in the sample, and a polarization angle from a uniform distribution. This yields model Stokes $Q_0$ and $U_0$ to which Gaussian noise is added, with standard deviation equal to noise in Q/U images local to the source. This approach allows us to generate model $p_i$ that include the range in flux density of the sample, the sample size, and the local noise level. We generate many (typically 100 to 3000) independent realizations of this model stack, eliminating variations due to sample size in the model. In our analysis, we use the variance of the median and the quartiles of $p_i$ derived from independent realizations of the stack to determine statistical errors for these parameters that automatically include the effect of sample size.
    
    Figure~\ref{fig:p_cumdist_2panel} shows examples of model cumulative distributions of $p_i$, each representing the average of 100 Monte Carlo realizations of a stack. The only difference between the model curves in each panel is the assumed distribution of fractional polarization $\Pi_0$. For this example we used distributions that apply to our samples as described below: a Gaussian (blue curve and cyan curve), the distribution in Equation~\ref{eq:compact_source_dist} (dark blue curve) and the distribution in Equation~\ref{eq:double_source_dist} (red curve). The colors of the curves refer to the samples to which the distribution functions are found to apply in the analysis that follows. For example, dark blue refers to the \textit{Single a} sample as in Figures~\ref{fig:compact_selection}, \ref{fig:NVSS_souce_counts_bar_graph} and subsequent Figure~\ref{fig:quartile_ratio-fig}, etc. Each distribution was scaled to a median $\Pi_{0,\rm med}$ to fit the median $p_{\rm med}$ of the data. We apply one distribution function of $\Pi_0$ to all flux bins of a sample, allowing only the median $\Pi_{0,\rm med}$ to vary with flux density.
    
    The model cumulative distributions shown in Figure~\ref{fig:p_cumdist_2panel} all intersect the cumulative distribution of the data at the median because they represent fits to the median polarized intensity of the data. They may, however, deviate elsewhere from the cumulative distribution of the data. Deviations may also occur because of the finite sample size, and exceptional data values related to image errors and confusing sources that populate the extreme wing of the distribution. We use median stacking to minimize the effect of outliers. With the same argument, we use the quartiles $p_{25\%}$ and $p_{75\%}$ to quantify differences between the model and observed cumulative distributions in a way that is robust against outliers. Specifically, the ratio of these quartiles gives us one number that, in conjunction with the requirement to fit the median, quantifies agreement between the Monte Carlo distributions and the data (Figure~\ref{fig:p_cumdist_2panel}). \citet{stil_keller_george_taylor_2014} pointed out that the effect of the distribution function of $\Pi_0$ is much smaller than the total polarization bias correction, especially at a high signal to noise ratio.  The effect of the assumed distribution of $\Pi_0$ changes the derived median fractional polarization by an amount that is more than the statistical errors but less than the differences between samples. We return to this in Section~\ref{sec:discussion}, after we present our results.
    
    It is always possible to reproduce the median polarized intensity of the sample assuming a distribution of $\Pi_0$ for which the median $\Pi_{0,\rm med}$ is a fitting parameter, by scaling the distribution function. However, an arbitrary distribution will not reproduce the quartile ratio as a function of flux density, as shown by \citet{stil_keller_george_taylor_2014}. The requirement to match the quartile ratio can be met by adjusting the distribution function of $\Pi_0$ in the Monte Carlo simulations. We do not expect to retrieve the actual distribution function, but we can thus make a parameterization of the distribution of $\Pi_0$ that is consistent with the data. The main point is to define the number of sources in the sample with stronger polarized signal in relation to those with weaker polarized signal in a continuous way. To do this, we can afford some freedom in the parameters introduced in the description of the distribution function.
    
    Figure~\ref{fig:quartile_ratio-fig} shows quartile ratios as a function of flux density for different samples.  The error bars plotted on the data represent the standard deviation of independent Monte Carlo realizations that adopt the fitting $\Pi_{0,\rm med}$ and model distribution. A single distribution function reproduces the quartile ratios of a sample as a function of flux density reasonably well. Also shown are dashed curves that represent the quartile ratios of fits with distributions from other samples. These illustrate the statistical significance with which some distributions are rejected by the quartile ratio test.
    
    A Gaussian distribution of $\Pi_0$ reproduces the quartile ratios for samples \textit{Single b} and \textit{Single c} reasonable well (Figure~\ref{fig:quartile_ratio-fig}, top right and bottom left). However, assuming a Gaussian distribution of $\Pi_0$ that matches $p_{\rm med}$ for the most compact sample \textit{Single a}, results in the blue dashed curve in the top left corner of Figure~\ref{fig:quartile_ratio-fig}. A $\Pi_0$ distribution with a much stronger tail is required to reproduce the quartile ratios observed for the most compact sources. To this end, we used $\Pi_0 \ge 0$ distributions of the form 
    \begin{equation}
    f_1(\Pi_0) \sim {1 \over A + \Pi_0^\gamma},
    \label{eq:compact_source_dist}
    \end{equation}
    with $A = 1.95\%$ and $\gamma = 1.8$. This distribution yields a median fractional polarization of $2.23\%$, which can be adjusted by scaling the values drawn at random by the desired factor. 
    
    The \textit{Double} samples are also not well represented by a Gaussian distribution of $\Pi_0$ (blue dashed curve in the bottom right panel of Figure~\ref{fig:quartile_ratio-fig}). Here, a Gaussian distribution yields a quartile ratio that is too large. The $\Pi_0$ distribution that matches the most compact sample provides an even worse fit than the Gaussian, as shown in Figure~\ref{fig:p_cumdist_2panel} (bottom panel and the dark blue dashed curve in the bottom right panel of Figure~\ref{fig:quartile_ratio-fig}. Here, a distribution that tends more toward a uniform distribution is required. After some experimentation, a Gaussian with a positive mean value, discarding any negative values, was selected. The distribution function, for $\Pi_0 \ge 0$,
    \begin{equation}
        f_2(\Pi_0) \sim \exp{ \bigl[-\onehalf(\Pi_0-0.7)^2\bigr]},
        \label{eq:double_source_dist}
    \end{equation}
    yields a median fractional polarization $1.01\%$. The function was subsequently scaled along the $\Pi_0$ axis to adjust the median. The data do not require a distribution with a local minimum at $\Pi_0 = 0$, but there is no a priori reason to rule out such a distribution either. Considering the modest size of the double samples, this function proved adequate.
    
    \begin{figure*}[t!]
        \centering
        \includegraphics[width=0.95\linewidth]{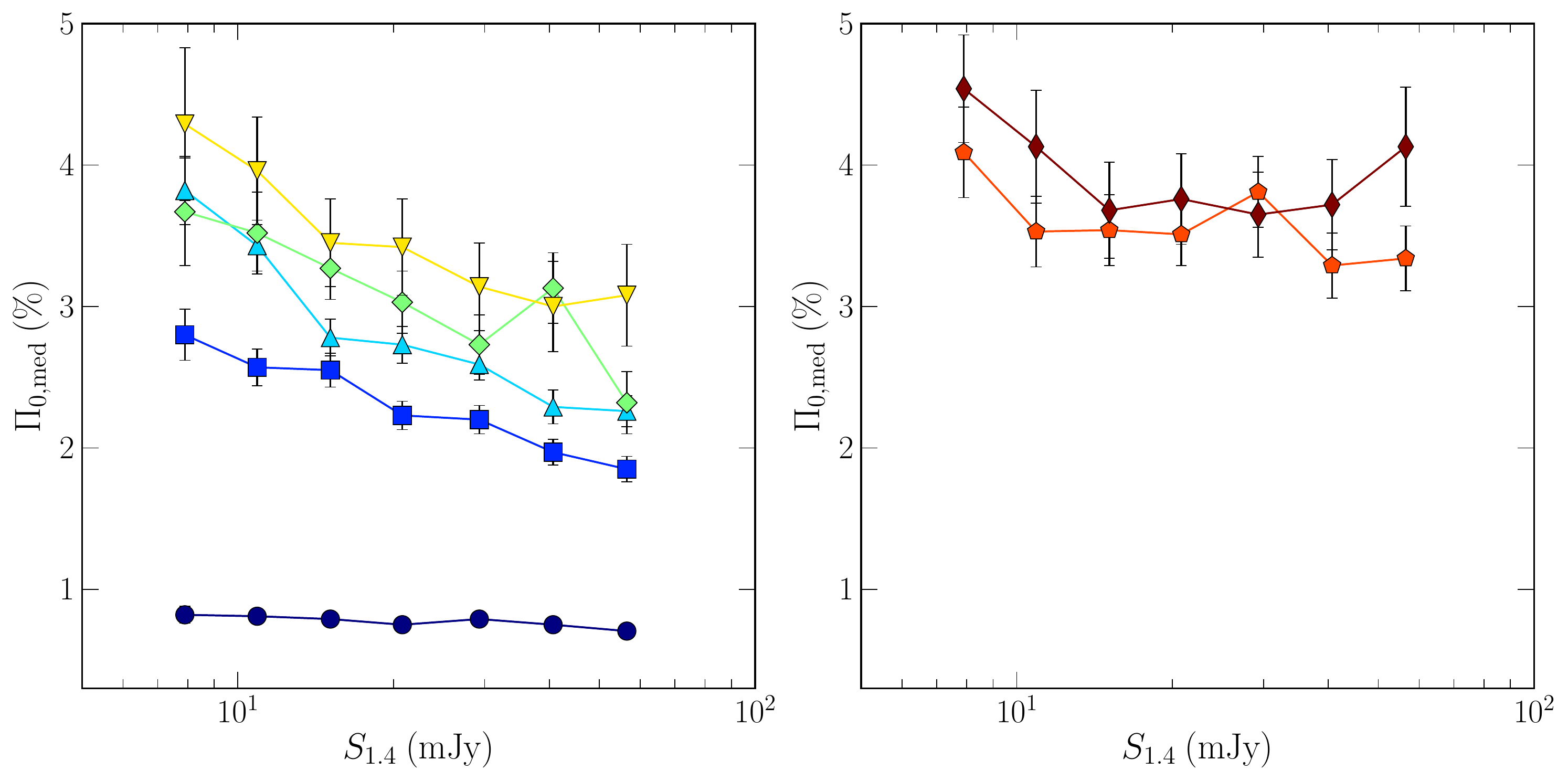}
        \caption{Fractional polarization as a function of flux density from stacking NVSS data. The left panel shows the \textit{Single} sources while the right panel shows the \textit{Double} sources. Each group of symbols shows samples selected of the same compactness $\mathcal{C}$: dark blue circles: \textit{Single a}; blue squares: \textit{Single b}; cyan triangles: \textit{Single c}; green diamonds: \textit{Single d}; yellow inverted triangles: \textit{Single e}. Right panel: orange pentagons: \textit{Double a}; red diamonds: \textit{Double b}. Error bars are derived from the Monte Carlo simulations and represent the 16.5 and 83.5 percentiles of distributions.}
        \label{fig:percent_pol}
    \end{figure*}
    
    It may at first glance appear strange that a single $\Pi_0$ distribution function leads to different quartile ratio curves in different panels of Figure~\ref{fig:quartile_ratio-fig}. These differences are most evident for the dark-blue curves in Figure~\ref{fig:quartile_ratio-fig}. The reason for this behaviour is the different signal to noise ratio of the sample that must be reproduced. In the construction of the quartile ratio curves, the $\Pi_0$ distribution is scaled to produce a median value of the observed polarized intensity after blending in the noise. For a sample with a higher median stacked polarized intensity, $p_{\rm med}$, the signal to noise ratio is higher, and therefore the quartile ratio is closer to the quartile ratio of the true distribution. The quartile ratio of the noise without signal is determined in turn by the range of noise values in the sample. This is very similar for each of the samples considered here, but we note that these samples are restricted to the sky coverage of the FIRST survey, which is not the same as the complete NVSS. 
    
    The quartile analysis can only be done for sufficiently large samples, in this case samples \textit{Single a, b, c}, and the combined {\it Double} sample. For the samples \textit {Single d} and \textit{Single e} we applied a Gaussian distribution without the quartile ratio test. If the true distribution is different, a small systematic error in true median fractional polarization ensues. We can estimate the size of this systematic error by repeating the analysis with different distributions. Adopting the distribution function for the most compact sample in the Monte Carlo simulations for the extended \textit{Sample 5c} yields a median $2.98\% \pm 0.43$. Table~\ref{tab:results} lists $\Pi_{0,\rm med} = 2.78 \% \pm 0.13\%$. The systematic difference in the derived true median $\Pi_{0,\rm med}$ is smaller than the statistical error for this small sample (see also Section~\ref{sec:discussion}). The quoted uncertainty depends on the assumed distribution, because the median observed polarized intensity is a complicated function of the distribution of the actual signal. In order to reproduce an $1 \sigma$ change in the observed median, the true median needs to be varied by a larger or smaller amount depending on the distribution function of $\Pi_0$.

    \begin{figure*}[t!]
        \centering
        \includegraphics[trim={0.2cm 0.1cm 0.1cm 0.1cm},clip,width=0.85\linewidth]{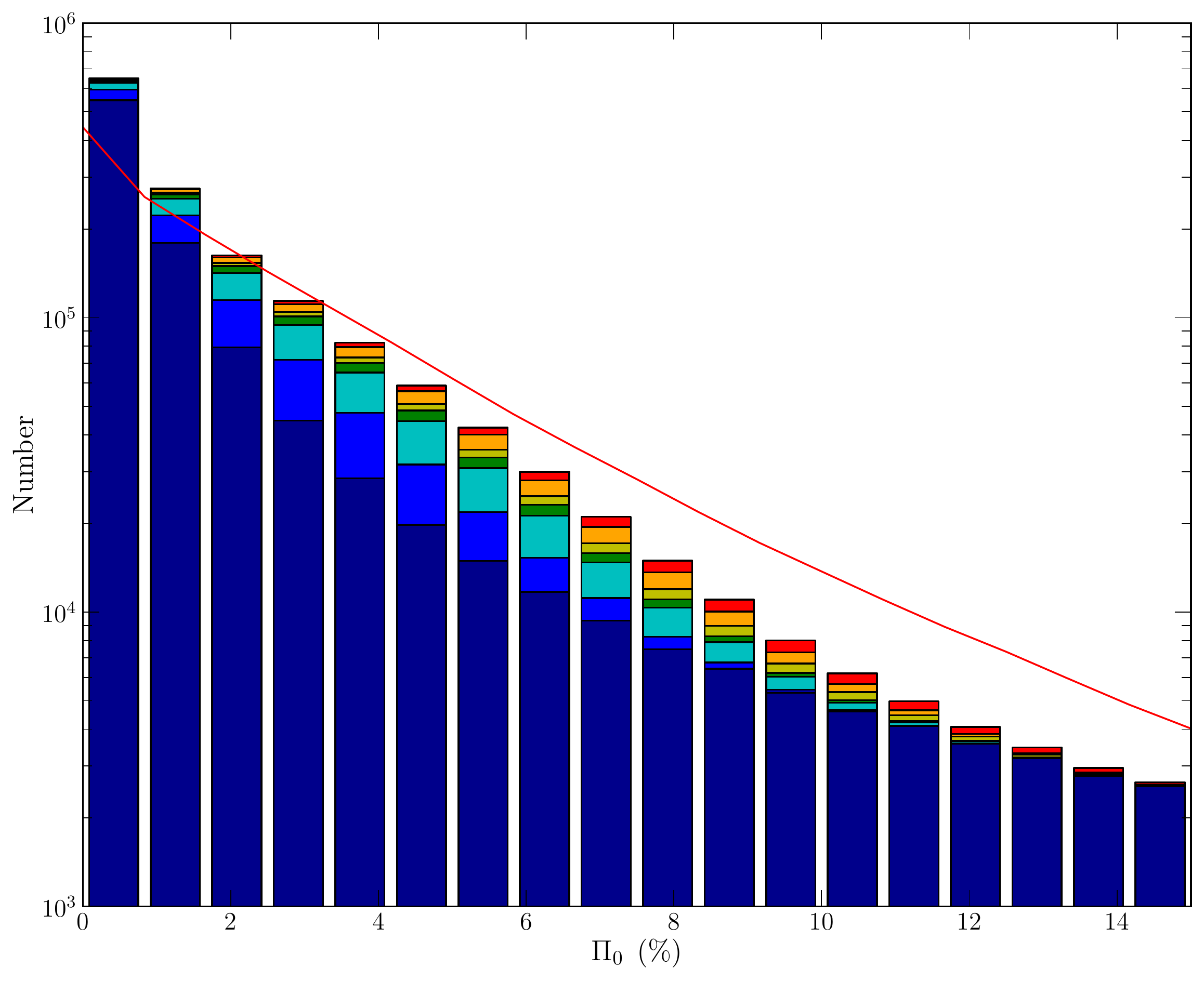}
        \caption{Monte Carlo realization of the distribution of $\Pi_0$ of the samples in this paper for flux density $50$ to $70$ mJy compared with a realization of the empirical distribution of \citet{Beck_Gaensler_2004} fitted to NVSS sources brighter than 100 mJy (red curve). The colour coding is the same as in Figure~\ref{fig:NVSS_souce_counts_bar_graph}. The sample size is normalized to $10^6$ sources in the most compact sample and sample size ratios according to the numbers in Table~\ref{tab:results}. The extreme tail of the distribution is not shown, as it is not well constrained.}
        \label{fig:Fracpol_simdist_fluxbin_1}
    \end{figure*}

    \subsection{$\Pi_{0,\rm med}$ as a function of size and flux density}    

    Figure \ref{fig:percent_pol} shows $\Pi_{0,\rm med}$ as a function of stacked 1.4 GHz flux density. Each data point represents a stacked sample with its own Monte Carlo analysis. The left panel shows data for the \textit{Single} samples, while the right panel shows results for the \textit{Double} samples. The most compact sources, represented by the dark blue circles, are the least polarized with median fractional polarization $\Pi_{0,\rm med} \lesssim 1.0\%$. This is less than half of the median fractional polarization of a complete sample of radio sources with the same flux density, which is significant considering that the most compact sample represents $\sim 70\%$ of our \textit{Single} sources, and at least a third of all NVSS sources (see Figure~\ref{fig:NVSS_souce_counts_bar_graph}). The next compactness bin ($1.15 \leq \mathcal{C} \leq 2.0$) is significantly more polarized with $\Pi_{0,\rm med} \sim 2.3\%$, which is higher for fainter samples.

    \citet{Cotton_2003} and \citet{Fanti_2004} reported an abrupt drop in 1.4 GHz fractional polarization of CSS and GPS sources compiled by \citet{Fanti_2001} at a size scale of $12\ \rm kpc$ (adjusted for cosmology). The scale at which this depolarization occurs is smaller at shorter wavelengths, indicating Faraday depolarization as a likely cause \citep{Fanti_2004}. The abrupt drop in fractional polarization from a few percent to less than $1\%$ is very similar to our present result, especially when comparing the 1.4 GHz polarization of all sources with size less than $\sim 40$ kpc (adjusted for cosmology) in \citet{Fanti_2004} with our samples. There are some differences. We make no selection on spectral index. The angular size derived from VLASS stacking (Section~\ref{sec:data_and_methods}) of our \textit{ Single a} sample is only slightly larger, about $15$ kpc. We also see a more gradual but statistically significant increase for more extended sources. Our samples are about an order of magnitude fainter than the sample of \citet{Fanti_2001}, and consequently the sample size is much larger. We will return to this comparison in Section~\ref{sec:discussion}. In the next section, we will see that the low-frequency spectrum of our samples does not turn over as CSS spectra tend to do, which is more in line with the recent sample of galaxy scale jets presented by \citet{Webster_2020}.
    
    The median fractional polarization continues to increase to $\Pi_{0,\rm med} \sim 3.2\%$ as $\mathcal{C}$ increases. These samples also show a tendency for higher fractional polarization in fainter sources, with the caveat that this trend is somewhat dependent on the $\Pi_0$ distribution, which is not well constrained for small samples (see Section~\ref{sec:discussion}). The right panel of Figure \ref{fig:percent_pol} shows the two \textit{Double} samples.  These are the most polarized of all our samples with $\Pi_{0,\rm med} \sim 4\%$, approximately independent of (total) flux density. 
    
    Figure~\ref{fig:Fracpol_simdist_fluxbin_1} shows a histogram of a Monte Carlo realization of the $\Pi_0$ distributions of all sub-samples for flux bin 1, normalized to $10^6$ sources in the most compact sample. The histogram is truncated at $15\%$ polarization to improve legibility. Extended sources are under-represented at the low and over-represented at the intermediate fractional polarization, relative to their part of the entire sample. The median fractional polarization of the combined sample is $1.16\%$, which is smaller than the $\sim 2\%$ median of NVSS sources with the same flux density. The red curve in Figure~\ref{fig:Fracpol_simdist_fluxbin_1} shows, for the same total sample size, the empirical $\Pi_0$ distribution of NVSS sources brighter than 100 mJy by \citet{Beck_Gaensler_2004} with 6\% unpolarized sources, which yields a median $1.89\%$. The distributions are superficially similar, but the accumulation of our samples is significantly skewed to lower $\Pi_0$. 
    
    We attribute the difference to our sample selection, which is likely very effective identifying single compact sources, but not designed to be complete for resolved sources. In particular, our requirement of positional correspondence within $3\arcsec$ between the FIRST and NVSS surveys is small compared with the source size of our more extended single source samples. Figure~\ref{fig:doubles_selection} suggests that allowing for more extended source components and more unequal fluxes could increase the sample of doubles by a factor of a few, at the risk of introducing more chance alignments and perhaps image artifacts. Furthermore, the angular size of the \textit{Single a} sample is well below the median angular size of sources at mJy flux densities (see also Figure~\ref{fig:extended_mosaic_1} to Figure~\ref{fig:unused_sources_mosaic}), but the \textit{Single a} sample represents $\sim 65\%$ of the cumulative sample. It is in line with our results that we find a distribution that is skewed to low polarization as our samples are more representative of the compact source population. Rejected sources as shown in Figure~\ref{fig:unused_sources_mosaic} may be more polarized, in line with the results of \citet{Rudnick_Owen_2014}.
    
    The median fractional polarization of bright radio sources in the NVSS is $\sim 2\%$, with a modest tendency of faint sources to be more polarized. We find that this dependence of fractional polarization on flux density is correlated with angular size. \citet{2015aska.confE.112S} presented results from stacking samples of radio sources selected by spectral index. These authors found that flat spectrum sources have a low fractional polarization consistent with the \textit{Single a} sample. However, the \textit{Single a} sample is too large to be made up of only flat spectrum sources. \citet{2015aska.confE.112S} found that steep spectrum sources had a consistent $\sim 2\%$ median fractional polarization, while sources with an intermediate spectral index showed a strong relation of fractional polarization with flux density. It is interesting that two completely different selections of sub-samples of radio sources reveal a similar pattern. To this end, we investigate also the spectral index of our samples through median stacking of total intensity in the NVSS and three other surveys with comparable angular resolution.

    \begin{figure*}[t!]
        \centering
        \includegraphics[width=0.8\linewidth]{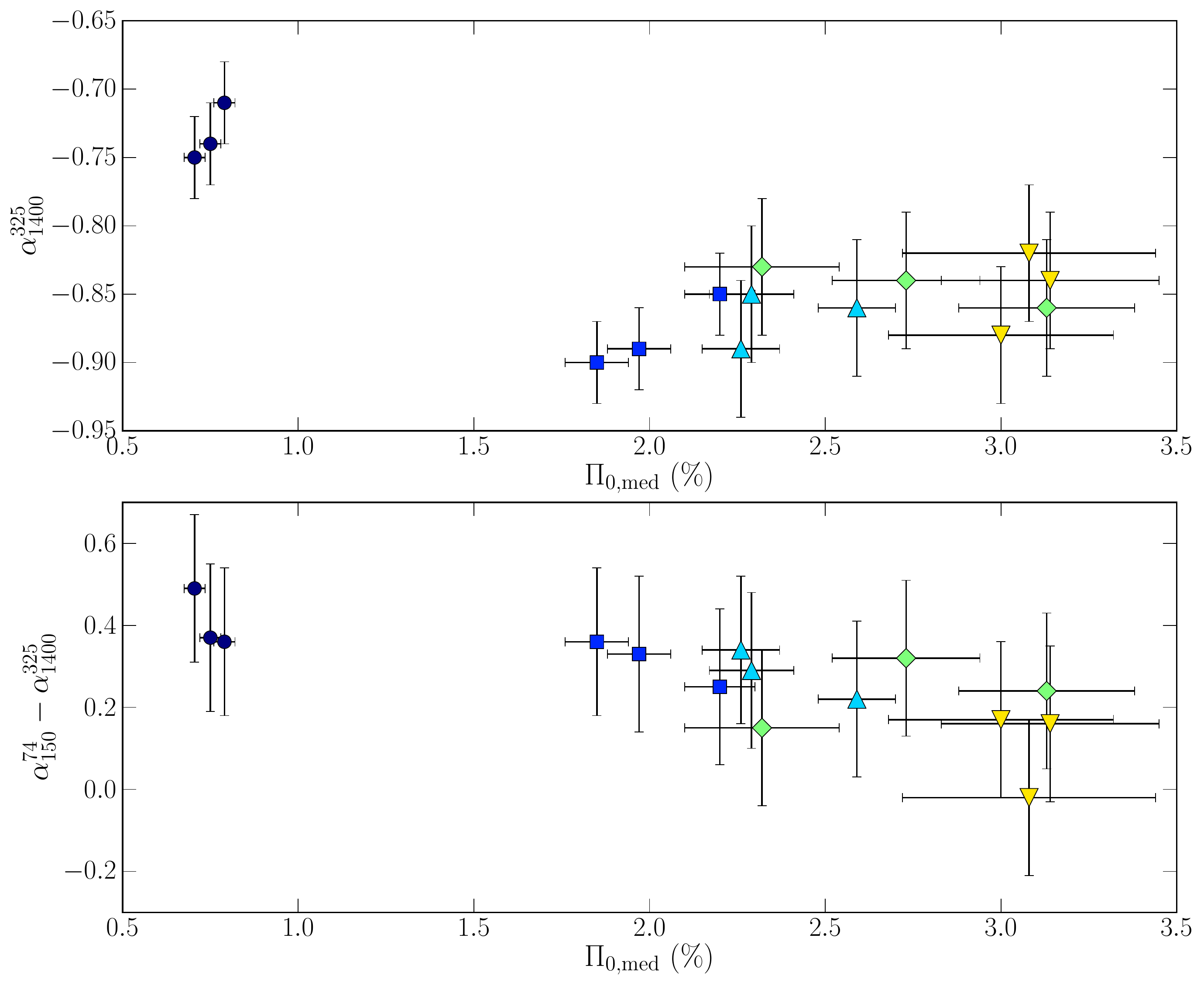}
        \caption{Results of spectral analysis for single sources. The top panel shows WENSS-NVSS spectral indices as a function of fractional polarization for \textit{Single a} through \textit{Single e} samples. The bottom panel shows the difference of WENSS-NVSS spectral index and the TGSS-VLSS spectral index as a function of fractional polarization. We only include the upper 3 flux bins because the lower 4 were deemed problematic for the low-frequency Stokes I stacking (see note in Table \ref{tab:results}).}
        \label{fig:Pi_vs_alpha}
    \end{figure*}
    
    \begin{figure}[b!]
        \centering
        \includegraphics[trim={0.25cm 0.2cm 0.25cm 0.25cm},clip,width=\linewidth]{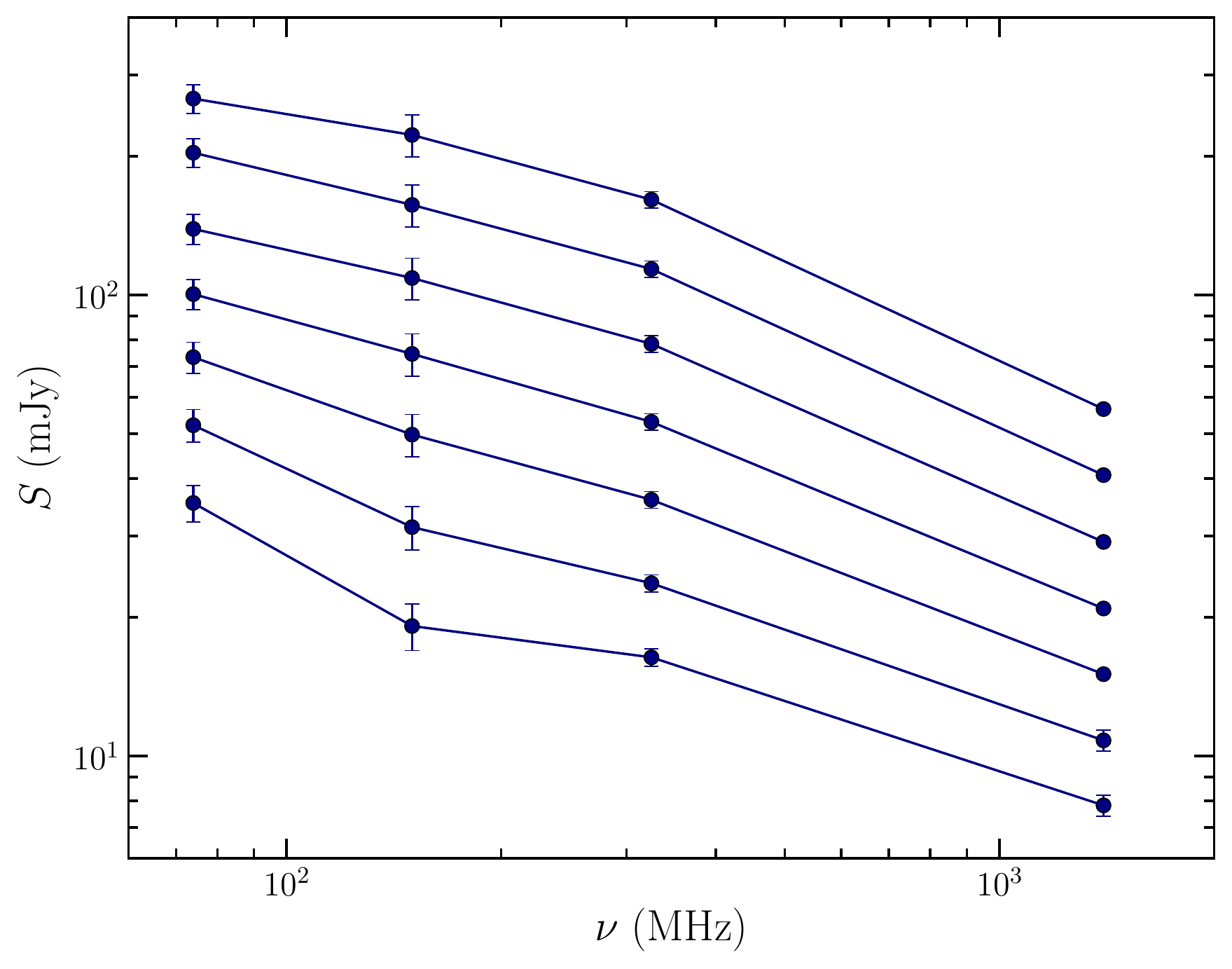}
        \caption{Flux density plotted as a function of frequency for all seven \textit{Single a} samples listed in Table \ref{tab:results}. The error bars represent the measured background noise in the median stacked images. The slope is the spectral index $\alpha$.}
        \label{fig:flux_freq_plot}
    \end{figure}
    
    \subsection{Spectral index from median stacking}
    
    We included each of NVSS, WENSS, TGSS, and VLSS to derive spectral indices\footnote{We convert literature values to our sign convention for spectral index $\alpha$ where necessary.}  $\alpha_{1400}^{325}$, $\alpha_{325}^{150}$, and $\alpha_{150}^{74}$ from the median stacked Stokes I images. The errors in the spectral indices were derived from the noise in the images using standard error propagation. This does not take into account that the samples were drawn from a population of sources with an intrinsic range of spectral indices. The accuracy with which one can derive the sample median from a sample of finite size is limited. The intrinsic range of spectral indices of the samples cannot be determined from the available data. An estimate can be obtained by simulating the sample median of the spectral index distribution of bright sources published by \cite{DeBreuck_2000}. The median spectral index of their distribution is $-0.785$. Assuming noiseless data, the uncertainty on the median is 0.023 for a sample size of 120, 0.017 for a sample size of 200, 0.010 for sample size of 500, and 0.0033 for sample size of 5000. The accuracy of the spectral indices in Table \ref{tab:results} is therefore limited by sample size and survey sensitivity.
    
    Some care is required when comparing the spectral indices in Table \ref{tab:results} to the literature. The approach we take includes all sources in our sample selected by NVSS flux density, regardless of whether they are detected in the other surveys. This is substantially different from most of the spectral index studies found in the literature, which depend on a cut-off imposed by the detection threshold in both of the surveys \citep [e.g.][]{Williams_2013}. This does not occur for us; we rely on the spectral index from the median flux density to trace the representative value of $\alpha$. \cite{Marvil_2015} derived spectral indices in the same manner as we do (mean of $\alpha_{325}^{74}=-0.45$). However, their sample consists of bright spiral galaxies and not AGN. 
    
    Upon inspection of Table \ref{tab:results}, there is a trend for the spectrum of all our samples to become flatter ($\alpha$ increases) with fainter flux density, although not all by the same amount. The effect is strongest for the doubles, with a difference of $\Delta \alpha_{1400}^{325} = +0.25$, followed by the compact sources ($\Delta\alpha_{1400}^{325} = +0.22$), and is smallest for the extended single sources ($\Delta \alpha_{1400}^{325} = +0.16$). 
    
    Figure \ref{fig:Pi_vs_alpha} shows the spectral index ($\alpha_{1400}^{325}$ and spectral curvature $\alpha_{150}^{74}- \alpha_{1400}^{325}$) from stacking in relation to their median fractional polarization for the three brightest flux bins. In the top panel, we not only see the divide in $\Pi_{0,\rm med}$ between compact and extended sources, as in Figure \ref{fig:percent_pol}, we also see a consistent and significant difference in $\alpha_{1400}^{325}$ between the compact and extended sources. In the bottom panel, the difference $\alpha_{150}^{74} - \alpha_{1400}^{325}$ shows the curvature of the spectrum, where lower polarization is associated with a flatter low-frequency spectrum. The Pearson correlation coefficient is $-0.76$. \citet{Webster_2020} showed a similar diagram for sources that they resolved. These authors probed similar size scales and found similar spectral indices, although the scatter in their relation is large. 
    
    Figure \ref{fig:flux_freq_plot} shows the spectral energy distributions (SEDs) of the \textit{Single a} samples ($\mathcal{C}\leq 1.15$) on a logarithmic scale. We find a median spectral index of $\alpha_{1400}^{325} \sim -0.70$ and $\alpha_{325}^{150}\sim -0.50$, with a tendency to become flatter at fainter flux densities to $\alpha_{1400}^{325} \sim -0.60$ and $\alpha_{325}^{150}\sim -0.30$. There is a noticeable difference between the samples at frequencies of 150 MHz and 74 MHz. At brighter flux density, the low-frequency spectrum is very flat ($\alpha_{150}^{74}\sim -0.3$) and becomes very steep ($\alpha_{150}^{74}\sim -0.9$) at the faintest flux density. This effect appears in other stacked samples as well, and it is stronger for faint sources. It is an artifact of either VLSS or TGSS at low brightness levels. Differences in the flux calibration of surveys should have a proportional effect on sources of all brightness, so we do not attribute this problem to flux calibration standards. We can also discard confusion with faint sources, because such sources should be visible in the FIRST and VLASS images. For flux bins 4 to 7 ($6.6 \leq S_{1.4} \leq 25.5$ mJy), we don't consider $\alpha_{325}^{150}$ and $\alpha_{150}^{74}$ to be reliable. These values are listed between parentheses in Table~\ref{tab:results}.
    
    It is intriguing that there are differences in the change in $\alpha$ that are large compared to the uncertainty. It is difficult to attribute this trend to systematic errors because we see differences between samples of the same brightness, and many sources are well above the Stokes I noise level in all four surveys. We note that sample selection for sources in the 6.6 to 70.0 mJy 1.4 GHz flux density range is affected by detection thresholds in some all-sky surveys, and by small-number statistics in deep fields. There is some debate in the literature about a possible change of spectral index with flux density \citep[and references therein]{Randall_2012}. A number of papers claim that the spectrum of radio sources gets flatter as they get fainter \citep{Intema_2011,Mahony_2016}. For radio sources selected at 1.4 GHz, \cite{Owen_Morrison_2008} found that $\alpha_{1400}^{325}$ tends to flatten for sources with $S_{1.4}<10$ mJy, and angular sizes larger than $ 3\arcsec$. However, \cite{Owen_Morrison_2008} find that for the faintest flux densities ($\sim 0.5$ mJy) $\alpha_{1400}^{325}$ steepens again. This is in broad agreement with what we find in this paper. Other papers find no evidence of spectral index flattening at low flux densities \citep{Prandoni_2006,Ibar_2009,Mauch_2013}. 
    
    The values we find for $\alpha_{1400}^{325}$ are in agreement with \cite{Mauch_2013} who found a median spectral index of $\alpha_{1400}^{325}=-0.71$ for a sample of 5,263 sources detected in the NVSS and the H-ATLAS survey \citep{Herschel_ATLAS}. \cite{Bornancini_2010} analyzed the relationship between the spectral index and the environment of radio galaxies with hosts identified in the SDSS. \cite{Bornancini_2010} find that the mean spectral index for sources in galaxy clusters is $\alpha_{1400}^{325} = -0.65$, with steeper spectra associated with rich clusters, and no clear trend of flattening towards low frequencies. \cite{Randall_2012} found median spectral indices of $\alpha_{1400}^{843} = -0.775$, $\alpha_{1400}^{843} = -0.584$ and $\alpha_{1400}^{843} = -0.533$ for sources selected at $26 \leq S_{1.4} \leq 95.4$, $13.5 \leq S_{1.4} \leq 26$, and $6.4 \leq S_{1.4} \leq 9.8$ mJy. 
    
    Very compact radio sources become optically thick to their own synchrotron emission due to synchrotron self-absorption. A radio source that is opaque at low frequencies ($\sim100$ MHz) can be transparent at higher frequencies (i.e. 1.4 GHz). This may explain why the compact samples have a flatter low-frequency spectrum than the extended \textit{Single} and \textit{Double} samples. 

    We find that $\mathcal{C}>1.15$ sources have a steeper spectrum $\alpha_{1400}^{325} \approx -0.80$, the double sources having the highest spectral index of $\alpha_{1400}^{325} \approx -0.90$. Extended and double sources tend to have a steeper spectrum due to synchrotron ageing \citep[and references therein]{1985ApJ...291...52M}. \cite{Grant_2010} found a mean spectral index of $\alpha_{1420}^{325}=-0.77$ for 343 sources detected in the DRAO ELAIS N1 deep field. Furthermore, \cite{deGasperin_2018} found a mean spectral index of $\alpha_{1400}^{150} = -0.7870$ with standard deviation of $\sigma = 0.24$ for 503,647 sources in the NVSS and TGSS surveys. This is in agreement with \cite{Hurley-Walker_2017}, who found a median spectral index of $-0.78$ for a sample of 122,959 sources ($S_{200} > 160$ mJy). Although \cite{Hurley-Walker_2017} observed the sky from 72 to 231 MHz, the authors included estimates of the spectral indices over higher frequency ranges. \cite{Mauch_2003} found a median of $\alpha_{1400}^{843} = -0.83$ from cross-matching sources in the SUMSS and NVSS surveys. Recently, a study by \cite{Ishwara_Chandra_2020} examined the spectral indices of of 6,400 faint radio sources between 610 and 1400 MHz with a median flux density of $S_{610}=4.5$ mJy and found a mean $\alpha_{1400}^{610} = -0.85 \pm 0.05$. Compared to the literature, the reported values are very similar to what we find for extended and doubles sources, having typical spectral indices for sources selected at 1.4 GHz.

    At lower frequencies, our compact samples have flatter spectra, with spectral indices ranging from $\alpha_{325}^{150}=-0.47$ to $-0.37$. There is a significant difference between the literature and what we derived from the median stacked images. \cite{Williams_2013} found a mean spectral index of $\alpha_{327}^{153}= -0.84 \pm 0.02$ for 1,289 sources cross-matched with NVSS, WENSS and T-RaMiSu. \cite{Heald_2015} found a median $\alpha_{160}^{120} = -0.77$ for sources detected using the LOFAR High Band Antenna. \cite{Hurley-Walker_2017} also found a median $\alpha_{231}^{71} = -0.78$ for 245,470 sources, in agreement with \cite{Heald_2015}. Recall that in Figure \ref{fig:NVSS_souce_counts_bar_graph} $\sim 50\%$ of all NVSS sources present in the footprint of the FIRST survey did not show up in our samples. Our most compact samples (\textit{Single a}) consist of $33.8\%$ to $16.7\%$ of available sources, and only a combined $17.6\%$ to $7.5\%$ are extended and double sources. The difference between our results and those of the literature may be due to the restrictions imposed during sample selection (see Section~\ref{subsec:sample_Selection} and Figure~\ref{fig:unused_sources_mosaic}).

\section{Discussion} \label{sec:discussion}

\subsection{Technical considerations}
One of the most powerful applications of stacking is to compare the medians of samples selected from the same survey, but distinguished by an independent selection criterion that is related to a physical property of the targets. In this paper, we stack polarized intensity for samples of radio sources selected by total flux density and compactness. We find very significant differences in the (raw) stacked polarized intensity as a function of the compactness parameter $\mathcal{C}$. The trend shown in Figure~\ref{fig:sample_PI_vs_I_images} is found for all flux bins.

These differences cannot be attributed to data quality issues such as polarization purity, variation of noise with position, etc., because each sample probes hundreds to thousands of positions distributed throughout the survey area. The mean background of the stacked polarized intensity images (Figure 6) is very consistent between the samples, as one would expect for samples drawn at random positions throughout the survey. The noise about the median off-source position in the stacked polarized intensity image is proportional to the inverse square root of the sample size, as shown in Figure~\ref{fig:noise_sample_size}. This is why the stacked image of a small sample of bright sources appears noisier than the stacked image of a large  sample of faint sources (e.g. in Figure~\ref{fig:sample_PI_vs_I_images}). The effective noise level is as low as $3\ \rm \mu Jy$ beam$^{-1}$ for the largest samples. 

The polarized signal to noise ratio in the NVSS for a single 8 mJy source that is 1\% polarized is just 0.28. At this low signal to noise ratio, the observed polarized intensity is dominated by noise bias and depends only weakly on the real signal. Deriving the observed median fractional polarization from the stacked median polarized intensity requires Monte Carlo simulations that include a prescription of the distribution function of fractional polarization. This distribution function is constrained by the quartile ratio of the observed polarized intensities of the sample (Figure~\ref{fig:quartile_ratio-fig}). The quartile ratio is sufficiently sensitive to the shape of the $\Pi_0$ distribution to derive key attributes of the distribution function, such as non-Gaussianity for samples of order $10^3$ sources and larger. 

\begin{figure*}[t!]
    \centering
    \includegraphics[width=0.8\linewidth]{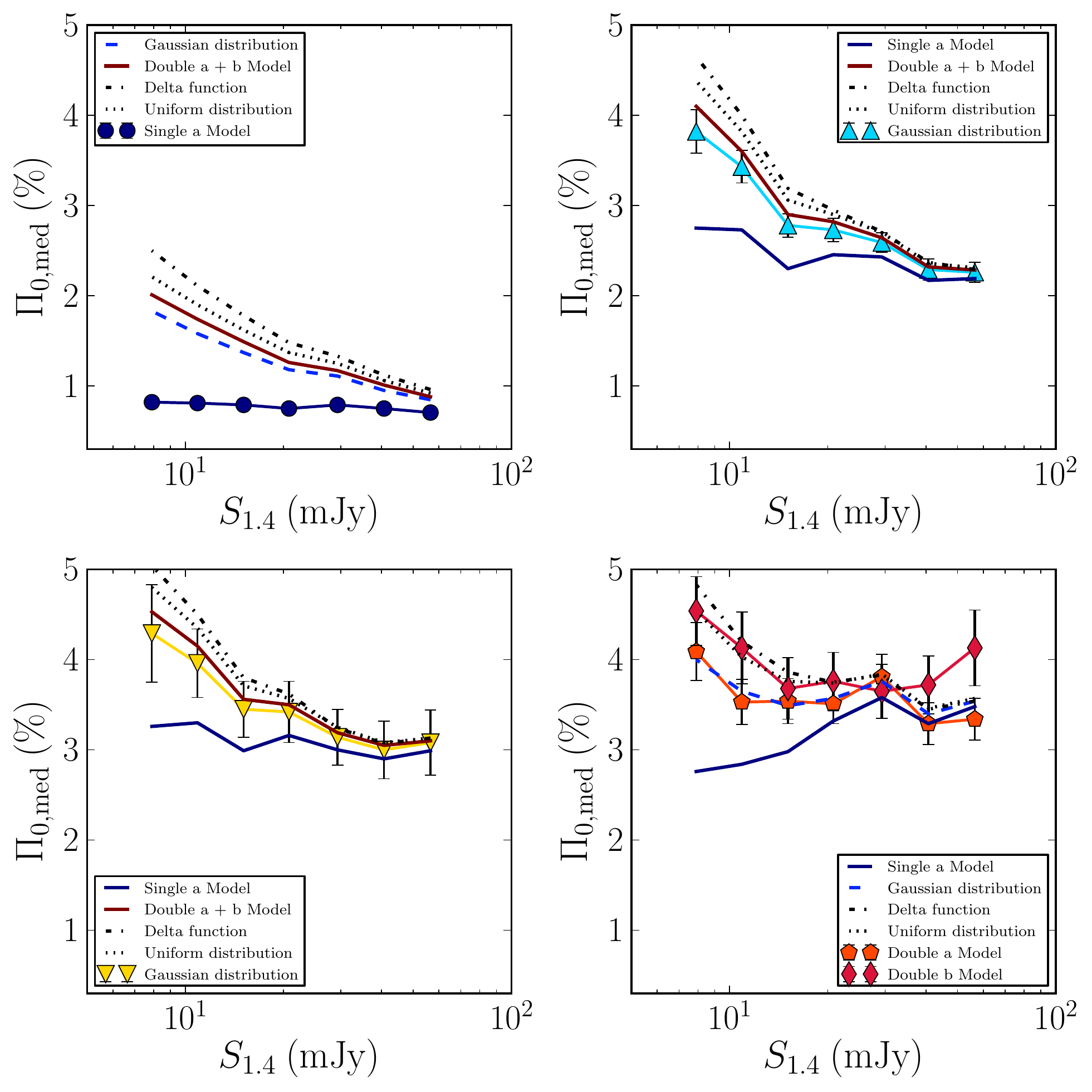}
    \caption{Reproduction of $\Pi_{0,\rm med}$ as a function of flux density for the samples \textit{Single a} (top left), \textit{Single c} (top right), \textit{Single e} (bottom left), and \textit{Double a} and \textit{Double b} (bottom right) shown in the same colors and symbols as in Figure~\ref{fig:percent_pol}. Also shown are curves of Monte Carlo simulations of the same stacks that reproduce $p_{\rm med}$, but assume a different distribution of $\Pi_0$. This illustrates the range of potential systematic errors if the quartile ratio analysis had not been considered. In fact, these curves are rejected by the quartile ratio test (Figure~\ref{fig:p_cumdist_2panel} and Figure~\ref{fig:quartile_ratio-fig}). For the brighter samples, our results are robust to the level of the statistical errors for a wide range of distributions. The effect of the distribution of $\Pi_0$ is larger for fainter samples \citep{stil_keller_george_taylor_2014}. The model curves represent the $\Pi_0$ distributions of other samples discussed in the text, as well as a delta function (all sources the same $\Pi_0$) and a uniform distribution. Colors of the curves refer to the samples for which they do fit the quartile ratio of the data, as in Figure~\ref{fig:quartile_ratio-fig}.}
    \label{fig:PI_vs_I_dist}
\end{figure*}

Incorporating the distribution of $\Pi_0$ leads to a modest adjustment of $\Pi_{0,\rm med}$ as illustrated in Figure~\ref{fig:PI_vs_I_dist}. Here we show fractional polarization as a function of flux density for the \textit{Single a}, \textit{Single c}, \textit{Single e}, and the \textit{Double} samples from Figure~\ref{fig:percent_pol}, along with curves that show the results that might have been obtained assuming various distributions of $\Pi_0$ without constraint by the quartile ratio analysis. The curves in each panel of Figure~\ref{fig:PI_vs_I_dist} differ only in the distribution of $\Pi_0$ assumed in the Monte Carlo analysis that fits $\Pi_{0,\rm med}$. The alternative distributions assumed in Figure~\ref{fig:PI_vs_I_dist} are almost all confidently rejected by the quartile ratio analysis (Figure~\ref{fig:p_cumdist_2panel} and Figure~\ref{fig:quartile_ratio-fig}), the uniform distribution applied to the \textit{Double} samples being a reasonably close match. Convergence of the curves for the brighter samples shows that the effect of the distribution reduces as the polarized signal to noise ratio of the median source increases. This must be so because $\Pi_{0,\rm med}$ is a fitting parameter and the effect of noise is reduced for brighter samples. 

At lower signal to noise ratio, the distribution of $\Pi_0$ matters because sources in the tail of the $\Pi_0$ distribution contribute disproportionately to higher $p_i$ \citep{stil_keller_george_taylor_2014}. Figure~\ref{fig:PI_vs_I_dist} illustrates that the $\Pi_0$ distribution has to be very different to have a noticeable effect, and such different distributions are confidently rejected by the quartile ratio constraint (Figure~\ref{fig:p_cumdist_2panel} and Figure~\ref{fig:quartile_ratio-fig}). In particular, Figure~\ref{fig:PI_vs_I_dist} underlines that the correlation of $\Pi_{0,\rm med}$ with flux density for the resolved \textit{Single} samples is not an effect of the assumed $\Pi_0$ distribution. The correlation would disappear if a skewed $\Pi_0 $ distribution like the one that applies to the \textit{Single a} was assumed. This distribution is strongly rejected because it does not reproduce the quartile ratio of the data for the other samples (Figure~\ref{fig:quartile_ratio-fig}). We conclude that the results shown in Figure~\ref{fig:percent_pol} are robust, even for samples where the quartile analysis could not be performed.

There is no significant difference in median 1.4 GHz Stokes I intensity in the stacked images of \textit{Single} samples with different compactness. The \textit{Double b} samples appear only $\sim 7\%$ fainter in median stacked total intensity. This confirms our implicit assumption that the $60\arcsec$ beam captures the integrated flux density of all samples. We measure comparable quantities for all samples in the polarization stacking as well as in the Stokes I stacking as a function of frequency, using surveys with $\sim 1'$ resolution. The panels in Figures~\ref{fig:extended_mosaic_1} to \ref{fig:unused_sources_mosaic} cover approximately the FWHM beam size of these surveys.

The combined samples represent approximately half of all NVSS sources within the footprint of the FIRST survey. That is a significant fraction, considering that some factors unrelated to the nature of a source can cause rejection by our sample selection. These include chance alignments, splitting a source into multiple catalog entries by the survey's source finder, and the effects of noise on $\mathcal{R}$ and $\mathcal{C}$, especially for the fainter samples. Some sources are too extended to be considered for stacking at a resolution of $1\arcmin$, or the brightness ratio of the lobes may be too dissimilar to be selected in our double sample, or the lobes are more diffuse. Also, large-amplitude variability may remove some sources from the samples with $\mathcal{R}\approx 1$. 

Our sample selection is quite restrictive on the double sources to provide a well-defined comparison sample for the \textit{Single} source samples. We recognize that among the sources that were not selected there are many extended sources and wider or asymmetric doubles (Figure~\ref{fig:unused_sources_mosaic}). The lower median fractional polarization of our accumulated sample when compared with the complete NVSS is in line with this finding and with the conclusions of this paper.

The high median fractional polarization of the double samples confirms that these samples represent mostly physical doubles as opposed to chance alignments. Any chance alignments of compact sources of the kind in our samples of single sources should result in integrated polarization that is less than the median of the single compact sources, because the polarization angles would not be correlated. On the contrary, the higher fractional polarization is consistent with aligned polarization vectors in the two parts. If not, even higher polarization is implied for one or both of the individual components. This can only be confirmed by future surveys, but it is likely that the two compact components of our double sources are lobes with hot spots at the termination points of the opposite jets as in FR II sources. Assuming symmetry in a statistical sense such that the 1.4 GHz polarization vectors in the doubles are aligned and equally polarized, we arrive at an interesting conclusion, because differential Faraday rotation between the opposite sides must not destroy this alignment. Each of the two sides experiences Faraday rotation by an amount $\Delta \theta = \phi \lambda^2$ with $\lambda^2 = 0.04585\ \rm m^2$ for the NVSS. If the Faraday depths $\phi$ of the medium in front of the two sides are different by just $22\ \rm rad\ m^{-2}$ differential Faraday rotation of one side relative to the other would amount to 1 rad at 1.4 GHz. The difference in rotation measure between the two sides should therefore be much less than $22\  \rm rad\ m^{-2}$ to maintain symmetry at 1.4 GHz. \citet{Vernstrom_2019} found the rms difference in RM of lobes of a sample of physical doubles to be only $4.6 \pm 1.1\ \rm rad\ m^{-2}$.  This is a remarkably tight limit for the differential Faraday rotation between the two sides of the double sources, considering the rms rotation measure of cluster cores is a few hundred $\rm rad\ m^{-2}$ \citep[e.g.][]{Clarke_2001}, and the standard deviation of RM across well-resolved radio galaxies is about $20\ \rm rad\ m^{-2}$ or larger \citep[e.g.][]{Taylor_1993,2008MNRAS.391..521L, Feain_2009,Anderson_2018,Banfield_2019}. This property makes resolved symmetric doubles prime background sources for experiments that require precision measurements of Faraday rotation \citep{2019arXiv190109074R}.

\subsection{Why a correlation between polarization and angular size?}

We find that the median fractional polarization of radio sources depends on angular size. The most compact sample, with angular size $\sim 1.8\arcsec$ has $\Pi_{0,\rm med} \lesssim 1\%$, although the derived $\Pi_0$ distribution is significantly skewed to higher fractional polarization. In the distribution shown in Figure~\ref{fig:Fracpol_simdist_fluxbin_1},  $\sim 40\%$ of compact sources is less than $0.5\%$ polarized, but $\sim 16\%$ of the compact sample is more than $3\%$ polarized. Similar fractions apply to other flux bins. These numbers suggest a sky density of $0.7$ per square degree for all 7 flux bins combined, or $3 \times 10^4$ in the sky. This serves to show that the low median fractional polarization of this sample does not exclude the presence of a significant fraction of polarized sources with angular size less than $2\arcsec$. The evidence for a skewed $\Pi_0$ distribution is the quartile ratio analysis presented in Figure~\ref{fig:p_cumdist_2panel} (top panel) and Figure~\ref{fig:quartile_ratio-fig} (top left panel). A peculiar consequence of this result is that the fraction of compact sources as a function of $\Pi_0$ initially decreases to a minimum around $\Pi_0 \approx 6\%$, and subsequently increases among the most polarized sources (Figure~\ref{fig:percent_pol}), notwithstanding the low $\Pi_{0,\rm med}$ of the sample.  

The second most compact sample, with angular size $\sim 2.9\arcsec$, is 2.5 to 3 times more highly polarized than the compact sample. The median fractional polarization of progressively more extended samples increases more slowly to $3\%$ to $4\%$ for sources with typical angular size $5\arcsec$ to $8\arcsec$ (40 to 70 kpc). Polarization of the \textit{Doubles} samples is consistent with this trend to somewhat larger angular scales.

Considering that the largest difference in $\Pi_{0, \rm med}$ is found between the two most compact samples, the question arises whether our most compact sample depolarizes at a smaller angular scale. After all, Figure~\ref{fig:vlass_mosaic_compact} shows a mixture of resolved and unresolved sources in the VLASS images of sample \textit{Single 5a}. To investigate this further, we evaluated the integrated flux density in cut-outs as shown in Figure~\ref{fig:vlass_mosaic_compact} and separated the sample by VLASS compactness at the ratio of the integrated flux density to the peak intensity equal to 1.4. The compact sub-sample has very low polarization $\Pi_{0,\rm med}= 0.5\%$. The more resolved sub-sample is more polarized with $\Pi_{0,\rm med} = 1.5\%$, but less so than any of the more extended samples (Figure~\ref{fig:percent_pol}). The polarization of the compact sub-sample is so low that our model for residual instrumental polarization \citep[$0.3\%$ rms of Stokes I leaking in each of Q and U, see also][]{Ma_2019} becomes a significant source of uncertainty. A complete analysis of VLASS compactness is better postponed until a VLASS source catalog becomes available, as this experiment was done with a subset of preliminary VLASS images. Still this result is more in line with a rapid increase in median fractional polarization across a range of angular scales from $\lesssim 2\arcsec$ to $\sim 5\arcsec$ and a more modest increase on even larger angular scales. This contrasts with the single precipitous drop on scales $\lesssim 1.5\arcsec$ ($\lesssim 12\ \rm kpc$), and smaller at shorter wavelengths, for bright CSS and GPS sources \citep{Cotton_2003}. \citet{Lamee_2016} also found a higher 1.4 GHz to 2.3 GHz depolarization factor for compact sources sources than for extended sources, although these authors considered flux densities and angular scales an order of magnitude larger than the present samples.

Could our compact source sample be dominated by similar faint CSS sources? \citet{Sadler_2016} suggested a population of faint CSS sources may have eluded detection, although the sample of CSS radio sources of \citet{Randall_2012} represents just a small fraction of the total population of faint radio sources. Our compact sample represents $68\%$ of our combined sample (including doubles), which amounts to $\gtrsim 34\%$ of all radio sources with flux density of tens of mJy. The lower limit indicates that some compact sources must have been rejected in our sample selection because of the proximity of an unrelated source, position errors, variability, etc. Because of this large percentage of the total population, our compact source sample should represent a larger diversity of less exceptional sources. We also see that $\Pi_{0,\rm med}$ increases on larger angular scales. Depolarization of CSS and GPS sources may just be the most extreme case of depolarization by Faraday rotation in the ISM of the host galaxy.

We focus the discussion on depolarization mechanisms related to the source and its immediate surroundings, as opposed to Faraday depolarization in unrelated objects along the line of sight. While intervening galaxies are a potential alternative cause of depolarization of radio sources with (sub) galactic size, the relatively high fractional polarization of our samples of symmetric double sources provides an argument that angular separation from the centre (i.e. the host galaxy), not solid angle of the radio emission, is the determining factor. The compactness of the components in the \textit{Doubles} samples is similar to that of the \textit{Single a} and \textit{Single b} samples. Also, there is at best weak evidence for depolarization associated with systems along the line of sight traced by optical Mg absorption lines \citep{Farnes_2014b, Chiaberge_2009,Kim_2016} or Ly-$\alpha$ absorbers \citep{Farnes_2017}.

Another reason to associate depolarization with the source itself is the correlation of $\Pi_{0,\rm med}$ with spectral index and spectral curvature (Figure~\ref{fig:Pi_vs_alpha}). \citet{2015aska.confE.112S} found a strong relation between $\Pi_{0,\rm med}$ and spectral index in the flux density range of this work. Flat spectrum sources ($\alpha^{325}_{1400} > -0.30$ by their definition) were found to be much less polarized than steep spectrum sources ($\alpha^{325}_{1400} < -0.75$). Flat and inverted spectrum sources are included in our samples, but the size of the compact sample is far too large to be made up of just flat-spectrum sources. We find that the spectral indices from median stacking indicate a slightly flatter spectrum for the compact sources, with not more than a hint of a break below $325\ \rm MHz$, but the range of all spectral indices $\alpha_{1400}^{325}$ from median stacking found here, qualifies as steep spectrum according to the most common definition \citep[e.g.][]{DeBreuck_2000}.

\citet{Farnes_2014a} found that flat spectrum sources and a subset of steep spectrum sources appear to have $\Pi_0$ that is nearly independent of wavelength. These authors raised opacity of different emitting regions of a compact source at different wavelength as a possible explanation why the median fractional polarization does not change above 1.4 GHz. At least qualitatively, our result appears consistent with this suggestion, in that our combined sample of compact sources, selected independently of spectral index, has similarly low $\Pi_{0,\rm med}$ as a separate sample of flat spectrum sources \citep{2015aska.confE.112S}. However, we do not expect optically thick synchrotron emission on scales $\sim 15\ \rm kpc$ in a majority of sources. The steep low-frequency spectrum implied by median stacking suggests that the fraction of sources whose flux density is dominated by compact components that become optically thick below 325 MHz, is small. 

The spectral index distribution of the sources for which \citet{Farnes_2014a} derived a polarization spectral index $\beta$ ($\Pi_0 \sim \lambda^\beta$), shows that steep spectrum sources are significantly under-represented in their final sample (compare their Figure 5 with their Figure 7). Steep spectrum sources that are depolarized on sub-galactic scales could be under-represented in the sample of sources with multi-wavelength polarimetry available in the literature compiled by \citet{Farnes_2014a}. At higher frequencies, a larger fraction of this sub-population could be polarized \citep{Fanti_2004,Rossetti_2008}, but they would be under-represented in a flux-limited sample at high frequencies because of their steep spectrum. \citet{Lamee_2016} found that $28\%$ of their steep spectrum sources ($24\%$ of their complete sample with $S_{2300}> 420$ mJy) were depolarized (1.4 GHz relative to 2.3 GHz) by a factor $\ge2$. A sub-sample selected to repolarize (low frequency polarization higher than high-frequency polarization) has a flat median spectral index $-0.1$, while sources with stronger depolarization had a steep spectral index $-0.9$. These authors also estimated they may have missed $\sim 10\%$ of their sample in the form of heavily depolarized sources. A better understanding of Faraday rotation effects in relation to source size and properties of the host galaxy will be key to unlocking the magnetic properties of AGN. 

If the low polarization of compact sources at 1.4 GHz arises from a flat spectrum optically thick core that is separate from a more extended steep-spectrum component, then this core should be visible in the VLASS images at 3 GHz. Figure~\ref{fig:vlass_mosaic_compact} shows first-look images of the VLASS survey for a random subset of sources in our most compact sample with angular resolution $2.5\arcsec$. Since the reference frequency of VLASS is a factor $\sim 2$ higher than that of the NVSS and FIRST, the sources span a wider range of flux density because of differences in spectral index. Also, flat spectrum components may be somewhat more prominent in the VLASS images. Still, the morphology of sources in the \textit{Single a} sample at $2.5\arcsec$ resolution is not that different from the more extended sources seen at $5\arcsec$ resolution in FIRST images (Compare Figure~\ref{fig:vlass_mosaic_compact} with Figure~\ref{fig:extended_mosaic_1} and Figure~\ref{fig:extended_mosaic_2}). We see elongated sources with symmetric lobes or edge-darkened morphology, not unlike the high-redshift FR I/FR II sample of \citet{Chiaberge_2009}. Opposite lobes of an edge-brightened source may blend at low resolution to appear as an edge-darkened source. \citet{Jimenez-Gallardo_2019} presented a sample of edge-brightened galaxy-scale radio sources that do not classify as GPS or CSS. Morphological classification of individual sources requires even higher resolution images. Roughly half of the \textit{Single a} sample in Figure~\ref{fig:vlass_mosaic_compact} contains a brighter, compact component, but by no means is the compact sample one of core-dominated sources. 

Linear polarization of optically thin synchrotron emission depends on the amount of order in the magnetic field in the source, and on differential Faraday rotation within the source or across the source. The structure of the source is important for detailed modeling of polarization as a function of wavelength. Currently, we have limited morphological information for our samples. Our \textit{Double} samples are most likely sources with FR II morphology. The structure of our \textit{Singles} samples is not so clear. CSS, GPS, and FR 0 \citep[e.g.][]{Sadler_2014,Baldi_2018} sources are most likely included in the \textit{Single a} samples. The steep low-frequency spectra of our \textit{Single a} samples suggest that these samples include a more diverse class of objects than just CSS and GPS. The more extended \textit{Single} samples may well contain a mixture of morphologies \citep[e.g.][]{Mingo_2019,Webster_2020}.

Our results imply that the size of the radio source in relation to the size of its host galaxy is an important parameter for its 1.4 GHz polarization. This would include not only young powerful radio galaxies identified with CSS and GPS sources, but also, at lower flux densities, less powerful radio galaxies subject to frustration by their host's interstellar medium, or those with intermittent activity. The largest scale to which we see an increase in $\Pi_{0,\rm med}$ ($5\arcsec$ to $8\arcsec$ or 40 to 70 kpc) is comparable to the gaseous halo of a galaxy, which suggests that the circumgalactic medium contributes to depolarization. 

Physically larger sources that project into a small angular size will display associated properties related to relativistic beaming, such as a bright core, flat spectrum, variability, and a strong Laing-Garrington effect. \citet{Anderson_2019} reported variability in Faraday rotation of pc scale components in BLAZARS over a 5-year period. In a sample of randomly oriented sources, the median source axis makes an angle of $60\degr$ with the line of sight. Relativistic beaming will increase the fraction of sources where the jet is well aligned with the line of sight, but the size of the compact source sample is too large to suggest that our samples differentiate more in projection effects than in physical size.

Next generation wide-area polarization surveys of the sky, POSSUM and VLASS, will provide direct measurements of the polarization of faint radio sources currently only accessible through stacking. These surveys will add information from Faraday rotation for large samples of sources, a key unknown for the samples that we investigated here. These surveys also open up new frontiers for stacking and other sub-threshold techniques as the weak polarized signal of most sources detectable in total intensity will remain below the detection threshold. 

Stacking these future surveys need not require huge resources in computing or data storage, as our stacking of VLASS quick-look images shows. Key to this is a reduction of scope that is tuned to available resources. A wide-area, but not necessarily all sky, subset of images with arcsecond resolution, while averaged over a consistent $\lambda^2$ range around a low frequency and a high frequency within the survey's frequency band, would create a small overhead in data archiving, while creating the opportunity to explore science with sensitivity that requires a sky survey with the next generation of telescopes, such as ngVLA and the Square Kilometre Array (SKA). 

To this end, one or two orders of magnitude increase in sensitivity by stacking samples up to a few $10^4$ in VLASS or POSSUM would be sufficient to probe polarization to the 1\% level at the detection limit of Stokes I, without exceeding scalability of the stacking code or the demonstrated domain of $\sqrt{N}$ improvement of the noise when stacking surveys (e.g. Figure~\ref{fig:noise_sample_size}). Polarization of sub-mJy sources will remain accessible only by stacking and narrow deep fields in the era of POSSUM and VLASS. Low-redshift star forming galaxies are expected to display very low 1.4 GHz polarization. This may become apparent as a sharp drop in median polarization in the flux density range where star forming galaxies begin to outnumber AGN. Advanced sample selection that includes data from sky surveys at other wavelengths, such as X-ray or infrared surveys will allow comparative studies of physically different sub-samples analogous to the work presented here.

Detection of individual sources in total intensity is required for the interpretation of the stacked polarized signal. The large samples available from future surveys will primarily aid in the quartile ratio analysis. The higher sensitivity of these surveys will also improve the sample selection based on $\mathcal{R}$ and $\mathcal{C}$.

\section{Conclusions} \label{sec:conclusions}
    In this paper we explore the integrated polarization of AGN as a function angular size in the flux density range 70 mJy down to 6.6 mJy by stacking polarized intensity from the NVSS survey. We investigate 7 compactness classes in 7 flux bins: 5 samples of \textit{Single} sources with increasing ratio of integrated flux density to peak intensity in the FIRST survey, and two samples of \textit{Double} sources. All samples are unresolved in the NVSS, allowing for mutual comparison of their median polarization properties.
    
    We find a strong dependence of median fractional polarization on angular size, especially for the samples with deconvolved sizes $1.8\arcsec$ and $2.9\arcsec$, as defined as the FWHM of the emission of the sample in mean-stacked VLASS images, deconvolved for the VLASS beam and NVSS position errors.  This confirms various earlier suggestions from deep fields in the literature of a relation between fractional polarization and angular size. Our much larger samples reveal a strong change in $\Pi_{0,\rm med}$ on angular scales $\lesssim 5\arcsec$ and a more gradual change on angular scales up to $\sim 8\arcsec$. The most compact sample ($70\%$ of all \textit{Single} sources) has $\Pi_{0,\rm med} \sim 0.8\%$, approximately one third the value of the complete population. More extended sources have a median fractional polarization in the range $2.2\%$ to $4\%$ that is also higher for fainter sources. Our samples of symmetric doubles are at the high end of this range. 
    
    The rapid change in fractional polarization in the angular size range $\sim 2\arcsec$ to $\sim 5\arcsec$  is suggestive of a relation with the size of the host galaxy, as previously noted by \citet{Rudnick_Owen_2014}, but it occurs on scales larger than depolarization reported for CSS sources by \citet{Cotton_2003}, suggesting that the circumgalactic medium plays a role. 
    
    Analysis of the derived distribution of fractional polarization shows that $16\%$ of our most compact sample is more than $3\%$ polarized with a sky density 0.7 per square degree. The skewed $\Pi_0$ distribution and large number of compact sources imply also that compact sources should be common amongst highly polarized sources, despite their low median polarization. This is good news for certain applications of a sky-wide Faraday Rotation Measure Grid, because the line of sight to these compact sources is much better determined than for the more extended samples.
 
    Symmetric double sources are the most polarized of our samples. This confirms that these are physical doubles, probably hot spots in sources with a FR II morphology. Symmetry of these sources also implies that differential Faraday rotation between the opposite sides of a symmetric double source is typically (much) less than $\sim 22\ \rm rad\ m^{-2}$. This makes these double sources excellent candidates to be top tier sources in the Rotation Measure Grid for probing magnetic fields along the line of sight. 
    
    Total intensity stacking of the samples in 4 separate arcminute-resolution radio surveys at 1.4 GHz (NVSS), 325 MHz (WENSS), 150 MHz (TGSS) and 74 MHz (VLSS) shows subtle but statistically significant differences in spectral properties between the samples. The spectral index derived from median-stacked total intensity between 325 MHz and 1.4 GHz is steep, but less so for the most compact sample ($\alpha_{1400}^{325} \approx -0.7$) and more so for the doubles $\alpha_{1400}^{325} \approx -0.9$, with the remaining samples in between. We find an anti-correlation between $\Pi_{0,\rm med}$ at 1.4 GHz and curvature of the low-frequency spectrum down to 74 MHz. This anti-correlation also provides a new argument that depolarization occurs in the source itself, as opposed to unrelated systems along the line of sight.

    \acknowledgments
    JMS acknowledges the support of the Natural Sciences and Engineering Research Council of Canada (NSERC), 2019-04848. The National Radio Astronomy Observatory is a facility of the National Science Foundation operated under cooperative agreement by Associated Universities, Inc. The authors thank the anonymous referee for insightful comments that helped improve the paper.

\bibliography{references}{}
\bibliographystyle{aasjournal}

\begin{deluxetable*}{lcccclcccc}
    \tablenum{1}
    \tablecaption{ Samples selected for Stacking.   \label{tab:samples_used}}
    \tablewidth{800pt}
    \tabletypesize{\scriptsize}
    \tablehead{
    \colhead{Sample ID} & 
    \colhead{Sample Size} & 
    \colhead{$r$} & 
    \colhead{$S_{\rm NVSS}$} & 
    \colhead{$\mathcal{C}$ Range} &
    \colhead{Sample ID} & 
    \colhead{Sample Size} & 
    \colhead{$r$} & 
    \colhead{$\rm S_{NVSS}$} & 
    \colhead{$\mathcal{C}$ Range} \\
    \colhead{} & \colhead{} & \colhead{($\arcsec$)} & \colhead{(mJy)} & 
    \colhead{} & \colhead{} & \colhead{} & \colhead{($\arcsec$)} & \colhead{(mJy)} &
    }
    \startdata
    Single 1a & $3634$ & $0-3$ & $50.0-70.0$ & $0.9-1.15$ & Single 2a & $4861$ & $0-3$ & $35.7-50.0$ & $0.9-1.15$ \\
    Single 1b & $717$ & $0-3$ & $50.0-70.0$ & $1.15-1.4$ & Single 2b & $952$ & $0-3$ & $35.7-50.0$ & $1.15-1.4$ \\
    Single 1c & $605$ & $0-3$ & $50.0-70.0$ & $1.4-1.9$ & Single 2c & $737$ & $0-3$ & $35.7-50.0$ & $1.4-1.9$ \\
    Single 1d & $180$ & $0-3$ & $50.0-70.0$ & $1.9-2.4$ & Single 2d & $235$ & $0-3$ & $35.7-50.0$ & $1.9-2.4$ \\
    Single 1e & $106$ & $0-3$ & $50.0-70.0$ & $2.4-5.0$ & Single 2e & $138$ & $0-3$ & $35.7-50.0$ & $2.4-5.0$ \\
    Double 1a & $189$  & $5-7$ & $50.0-70.0$ & $0.9-1.4$ & Single 2a & $245$ & $5-7$ & $35.7-50.0$ & $0.9-1.4$ \\
    Double 1b & $99$  & $7-9$ & $50.0-70.0$ & $0.9-1.4$ & Single 2b & $140$ & $7-9$ & $35.7-50.0$ & $0.9-1.4$ \\
    \hline
    Single 3a & $6204$ & $0-3$ & $25.5-35.0$ & $0.9-1.15$ & Single 4a & $7567$ & $0-3$ & $18.2-25.5$ & $0.9-1.15$ \\
    Single 3b & $1149$ & $0-3$ & $25.5-35.0$ & $1.15-1.4$ & Single 4b & $1315$ & $0-3$ & $18.2-25.5$ & $1.15-1.4$ \\
    Single 3c & $891$ & $0-3$ & $25.5-35.0$ & $1.4-1.9$ & Single 4c & $1035$ & $0-3$ & $18.2-25.5$ & $1.4-1.9$ \\
    Single 3d & $291$ & $0-3$ & $25.5-35.0$ & $1.9-2.4$ & Single 4d & $347$ & $0-3$ & $18.2-25.5$ & $1.9-2.4$ \\
    Single 3e & $168$ & $0-3$ & $25.5-35.0$ & $2.4-5.0$ & Single 4e & $215$ & $0-3$ & $18.2-25.5$ & $2.4-5.0$ \\
    Double 3a & $284$ & $5-7$ & $25.5-35.0$ & $0.9-1.4$ & Double 4a & $377$ & $5-7$ & $18.2-25.5$ & $0.9-1.4$ \\
    Double 3b & $157$ & $7-9$ & $25.5-35.0$ & $0.9-1.4$ & Double 4b & $192$ & $7-9$ & $18.2-25.5$ & $0.9-1.4$ \\
    \hline
    Single 5a & $8446$ & $0-3$ & $13.0-18.2$ & $0.9-1.15$ & Single 6a & $8474$ & $0-3$ & $9.3-13.0$ & $0.9-1.15$ \\
    Single 5b & $1425$ & $0-3$ & $13.0-18.2$ & $1.15-1.4$ & Single 6b & $1269$ & $0-3$ & $9.3-13.0$ & $1.15-1.4$ \\
    Single 5c & $1087$ & $0-3$ & $13.0-18.2$ & $1.4-1.9$ & Single 6c & $1017$ & $0-3$ & $9.3-13.0$ & $1.4-1.9$ \\
    Single 5d & $380$ & $0-3$ & $13.0-18.2$ & $1.9-2.4$ & Single 6d & $364$ & $0-3$ & $9.3-13.0$ & $1.9-2.4$ \\
    Single 5e & $230$ & $0-3$ & $13.0-18.2$ & $2.4-5.0$ & Single 6e & $253$ & $0-3$ & $9.3-13.0$ & $2.4-5.0$ \\
    Double 5a & $377$ & $5-7$ & $13.0-18.2$ & $0.9-1.4$ & Double 6a & $455$ & $5-7$ & $9.3-13.0$ & $0.9-1.4$ \\
    Double 5b & $185$ & $7-9$ & $13.0-18.2$ & $0.9-1.4$ & Double 6b & $228$ & $7-9$ & $9.3-13.0$ & $0.9-1.4$ \\
    \hline
    Single 7a & $7388$ & $0-3$ & $6.6-9.3$ & $0.9-1.15$ & & & & & \\
    Single 7b & $1100$ & $0-3$ & $6.6-9.3$ & $1.15-1.4$ & & & & & \\
    Single 7c & $781$ & $0-3$ & $6.6-9.3$ & $1.4-1.9$ & & & & & \\
    Single 7d & $330$ & $0-3$ & $6.6-9.3$ & $1.9-2.4$ & & & & & \\
    Single 7e & $226$ & $0-3$ & $6.6-9.3$ & $2.4-5.0$ & & & & & \\
    Double 7a & $538$ & $5-7$ & $6.6-9.3$ & $0.9-1.4$ & & & & & \\
    Double 7b & $384$ & $7-9$ & $6.6-9.3$ & $0.9-1.4$ & & & & & \\
    \enddata
    \tablecomments{Single sources were selected with $0.9 \leq \mathcal{R} \leq 1.1$ and $1.7 \leq \mathcal{R} \leq 2.3$ for double sources. The table is sorted in order of brightest to faintest $S_{NVSS}$.}
\end{deluxetable*}

\begin{deluxetable*}{cccccccc}
    \tablenum{2}
    \tablecaption{Results from the stacking, Monte-Carlo and spectral analysis.}
    \label{tab:results}
    \tablewidth{0pt}
    \tabletypesize{\scriptsize}
    \tablehead{
    \colhead{Sample ID} & 
    \colhead{$N_{\rm Stack}$} & 
    \colhead{${S}_{\rm NVSS}$} &
    \colhead{$\Pi_{0,\rm med}$} & 
    \colhead{$\alpha_{1400}^{325}$} & 
    \colhead{$\alpha_{325}^{150}$} & 
    \colhead{$\alpha_{150}^{74}$} & 
    \colhead{VLASS $\theta_{1/2}$} \\ 
    & & \colhead{(mJy)} & \colhead{median stack (\%)} & & & & ($\arcsec$)}
    \startdata
    Single 1a & $3571$ & $56.58$ & $0.705 \pm 0.03$ & $-0.75 \pm 0.03$ & $-0.47 \pm 0.16$ & $-0.26 \pm 0.18$ & $3.18$ \\    
    Single 1b & $704$  & $56.13$ & $1.850 \pm 0.09$ & $-0.90 \pm 0.03$ & $-0.64 \pm 0.16$ &  $-0.54 \pm 0.18$ & $3.93$ \\
    Single 1c & $594$  & $57.03$ & $2.260 \pm 0.11$ & $-0.89 \pm 0.05$ & $-0.70 \pm 0.16$ &  $-0.55 \pm 0.18$ & $5.48$ \\
    Single 1d & $174$ & $56.27$ & $2.320 \pm 0.22$ & $-0.83 \pm 0.05$ & $-0.70 \pm 0.16$ & $-0.68 \pm 0.19$ & $6.9$ \\
    Single 1e & $104$ & $56.32$ & $3.080 \pm 0.36$ & $-0.82 \pm 0.05$ & $-0.63 \pm 0.16$ & $-0.84 \pm 0.19$ & $8.6$ \\
    Double 1a & $184$ & $55.91$ & $3.340 \pm 0.23$ & $-0.97 \pm 0.05$ & $-0.69 \pm 0.16$ & $-0.62 \pm 0.19$ & \dots \\
    Double 1b & $94$  & $53.50$ & $4.130 \pm 0.42$ & $-0.91 \pm 0.05$ & $-0.82 \pm 0.17$ & $-0.71 \pm 0.19$ & \ldots \\ 
    \hline
    Single 2a & $4774$ & $40.69$ & $0.750 \pm 0.03$ & $-0.74 \pm 0.03$ & $-0.46 \pm 0.16$ & $-0.37 \pm 0.18$ & $3.28$ \\
    Single 2b & $934$ & $40.77$ & $1.970 \pm 0.09$ & $-0.89 \pm 0.03$ & $-0.62 \pm 0.16$ & $-0.56 \pm 0.19$ & $4.01$ \\
    Single 2c & $714$ & $40.63$ & $2.290 \pm 0.12$ & $-0.85 \pm 0.05$ & $-0.67 \pm 0.16$ & $-0.56 \pm 0.19$ & $5.48$ \\
    Single 2d & $231$ & $40.41$ & $3.130 \pm 0.25$ & $-0.86 \pm 0.05$ & $-0.63 \pm 0.16$ & $-0.62 \pm 0.19$ & $7.2$ \\
    Single 2e & $135$ & $41.59$ & $3.000 \pm 0.32$ & $-0.88 \pm 0.05$ & $-0.60 \pm 0.17$ & $-0.71 \pm 0.19$ & $8.5$ \\
    Double 2a & $241$ & $40.59$ & $3.290 \pm 0.23$ & $-0.92 \pm 0.05$ & $-0.77 \pm 0.17$ & $-0.59 \pm 0.20$ & \ldots \\
    Double 2b & $138$ & $39.31$ & $3.720 \pm 0.32$ & $-0.97 \pm 0.05$ & $-0.74 \pm 0.17$ & $-0.75 \pm 0.20$ & \ldots \\
    \hline
    Single 3a & $6101$ & $29.14$ & $0.790 \pm 0.03$ & $-0.71 \pm 0.03$ & $-0.37 \pm 0.16$ & $-0.35 \pm 0.18$ & $3.38$ \\
    Single 3b & $1134$ & $29.51$ & $2.200 \pm 0.10$ & $-0.85 \pm 0.03$ & $-0.62 \pm 0.16$ & $-0.60 \pm 0.19$ & $4.09$ \\
    Single 3c & $865$ & $29.42$ & $2.590 \pm 0.11$ & $-0.86 \pm 0.05$ & $-0.63 \pm 0.16$ & $-0.64 \pm 0.19$ & $5.70$ \\
    Single 3d & $285$ & $29.81$ & $2.730 \pm 0.21$ & $-0.84 \pm 0.05$ & $-0.62 \pm 0.17$ & $-0.52 \pm 0.19$ & $7.4$ \\
    Single 3e & $164$ & $29.43$ & $3.140 \pm 0.31$ & $-0.84 \pm 0.05$ & $-0.50 \pm 0.17$ & $-0.68 \pm 0.19$ & $8.6$ \\
    Double 3a & $275$ & $28.75$ & $3.810 \pm 0.25$ & $-0.93 \pm 0.05$ & $-0.71 \pm 0.17$ & $-0.82 \pm 0.20$ & \ldots \\
    Double 3b & $151$ & $27.69$ & $3.650 \pm 0.30$ & $-0.97 \pm 0.05$ & $-0.76 \pm 0.17$ & $-0.79 \pm 0.20$ & \ldots \\
    \hline
    Single 4a & $7450$ & $20.90$ & $0.750 \pm 0.03$ & $-0.67 \pm 0.03$ & ($-0.49 \pm 0.16$) & ($-0.43 \pm 0.19$) & $3.58$ \\
    Single 4b & $1294$ & $21.21$ & $2.230 \pm 0.10$ & $-0.82 \pm 0.03$ & ($-0.66 \pm 0.17$) & ($-0.19 \pm 0.19$) & $4.22$ \\
    Single 4c & $1008$ & $21.09$ & $2.730 \pm 0.13$ & $-0.84 \pm 0.05$ & ($-0.63 \pm 0.17$) & ($-0.74 \pm 0.19$) & $5.65$ \\
    Single 4d & $337$ & $20.78$ & $3.030 \pm 0.22$ & $-0.79 \pm 0.05$ &  ($-0.66 \pm 0.17$) & ($-0.64 \pm 0.20$) & $7.2$ \\
    Single 4e & $208$ & $21.04$ & $3.420 \pm 0.34$ & $-0.80 \pm 0.05$ & ($-0.46 \pm 0.17$) & ($-0.79 \pm 0.21$) & $8.6$ \\
    Double 4a & $367$ & $20.51$ & $3.510 \pm 0.22$ & $-0.89 \pm 0.05$ & ($-0.69 \pm 0.17$) & ($-0.86 \pm 0.20$) & \ldots \\
    Double 4b & $185$ & $19.78$ & $3.760 \pm 0.32$ & $-0.96 \pm 0.05$ & ($-0.60 \pm 0.17$) & ($-0.91 \pm 0.21$) & \ldots \\
    \hline
    Single 5a & $8202$ & $15.05$ & $0.790 \pm 0.04$ & $-0.63 \pm 0.03$ & ($-0.47 \pm 0.16$) & ($-0.56 \pm 0.19$) & $3.84$ \\                          
    Single 5b & $1386$ & $15.30$ & $2.550 \pm 0.12$ & $-0.80 \pm 0.03$ & ($-0.61 \pm 0.17$) & ($-0.82 \pm 0.20$) & $4.62$ \\
    Single 5c & $1052$ & $15.09$ & $2.780 \pm 0.13$ & $-0.75 \pm 0.05$ & ($-0.76 \pm 0.17$) & ($-0.63 \pm 0.21$) & $5.83$ \\
    Single 5d & $375$ & $14.98$ & $3.270 \pm 0.22$ & $-0.78 \pm 0.05$ &  ($-0.61 \pm 0.17$) & ($-0.91 \pm 0.22$) & $7.4$ \\
    Single 5e & $223$ & $14.99$ & $3.450 \pm 0.31$ & $-0.70 \pm 0.05$ & ($-0.62 \pm 0.17$) & ($-1.01 \pm 0.20$) & $9.0$ \\
    Double 5a & $366$ & $14.55$ & $3.540 \pm 0.25$ & $-0.85 \pm 0.03$ & ($-0.67 \pm 0.17$) & ($-0.97 \pm 0.22$) & \ldots \\
    Double 5b & $181$ & $14.03$ & $3.680 \pm 0.34$ & $-0.82 \pm 0.04$ & ($-0.59 \pm 0.17$) & ($-1.16 \pm 0.22$) & \ldots \\
    \hline
    Single 6a & $8257$ & $10.81$ & $0.810 \pm 0.05$ & $-0.57 \pm 0.05$ & ($-0.40 \pm 0.17$) & ($-0.73 \pm 0.20$) & $4.24$ \\
    Single 6b & $1227$ & $10.93$ & $2.570 \pm 0.13$ & $-0.69 \pm 0.05$ & ($-0.65 \pm 0.17$) & ($-0.89 \pm 0.21$) & $4.95$ \\
    Single 6c & $991$ & $10.89$ & $3.430 \pm 0.18$ & $-0.71 \pm 0.05$ & ($-0.73 \pm 0.17$) & ($-0.84 \pm 0.21$) & $5.99$ \\
    Single 6d & $357$ & $10.87$ & $3.520 \pm 0.29$ & $-0.68 \pm 0.05$ &  ($-0.73 \pm 0.18$) & ($-0.84 \pm 0.23$) & $7.3$ \\
    Single 6e & $246$ & $10.75$ & $3.960 \pm 0.38$ & $-0.70 \pm 0.05$ & ($-0.55 \pm 0.18$) & ($-0.95 \pm 0.24$) & $8.7$ \\
    Double 6a & $437$ & $10.46$ & $3.530 \pm 0.25$ & $-0.78 \pm 0.05$ & ($-0.64 \pm 0.17$) & ($-1.01 \pm 0.22$) & \ldots \\
    Double 6b & $320$ & $10.09$ & $4.130 \pm 0.40$ & $-0.72 \pm 0.06$ & ($-0.67 \pm 0.19$) & ($-1.22 \pm 0.25$) & \ldots \\
    \hline
    Single 7a & $7213$ & $7.81$ & $0.820 \pm 0.06$ & $-0.53 \pm 0.05$ & ($-0.26 \pm 0.18$) & ($-0.89 \pm 0.21$) & $4.46$ \\
    Single 7b & $1067$ & $7.89$ & $2.800 \pm 0.18$ & $-0.67 \pm 0.05$ & ($-0.56 \pm 0.18$) & ($-0.99 \pm 0.23$) & $5.03$ \\
    Single 7c & $757$ & $7.93$ & $3.820 \pm 0.24$ & $-0.66 \pm 0.05$ & ($-0.58 \pm 0.18$) & ($-1.04 \pm 0.24$) & $6.08$ \\
    Single 7d & $326$ & $7.90$ & $3.670 \pm 0.38$ & $-0.66 \pm 0.05$ &  ($-0.54 \pm 0.19$) & ($-1.08 \pm 0.26$) & $7.5$ \\
    Single 7e & $225$ & $7.86$ & $4.290 \pm 0.54$ & $-0.67 \pm 0.06$ & ($-0.48 \pm 0.19$) & ($-1.21 \pm 0.27$) & $7.9$ \\
    Double 7a & $523$ & $7.41$ & $4.090 \pm 0.32$ & $-0.71 \pm 0.05$ & ($-0.51 \pm 0.19$) & ($-1.02 \pm 0.26$) & \ldots \\
    Double 7b & $378$ & $7.28$ & $4.540 \pm 0.38$ & $-0.67 \pm 0.06$ & ($-0.58 \pm 0.20$) & ($-1.21 \pm 0.27$) & \ldots \\
    \enddata
    \tablecomments{Values for spectral index listed between parentheses are in the opinion of the authors affected by an unexplained systematic effect in the TGSS Redux or VLSS as explained in the text.}
\end{deluxetable*}

\end{document}